%% file: _main2.tex
\documentclass[a4paper,11pt]{article}
\usepackage{jcapmod}
\usepackage[a4paper,margin=2.5cm]{geometry}

\usepackage[utf8]{inputenc}
\usepackage[english]{babel}
\usepackage{graphicx}
\usepackage{bm}
\usepackage{physics} 
\usepackage{mathtools}
\usepackage{lipsum}
\usepackage{amssymb}
\usepackage{dsfont}
\usepackage{float}
\usepackage{booktabs}
\usepackage{array}
\usepackage{tabularx}
\usepackage{caption}

\usepackage{xcolor}
\usepackage{xspace}
\usepackage{ifthen}
\usepackage[font=scriptsize]{caption}
\graphicspath{{graphics}}

\usepackage[unicode=true,pdfusetitle,
 bookmarks=true,bookmarksnumbered=false,bookmarksopen=false,
 breaklinks=false,pdfborder={0 0 1},backref=false,colorlinks=true]
 {hyperref}
\hypersetup{
 linkcolor=blue, citecolor=cyan, urlcolor=red, filecolor=blue}

\usepackage{titlesec}
\titleformat{\subsubsection}
  {\normalfont\normalsize\itshape}{\thesubsubsection}{1em}{}

\usepackage{titling}
\newtheorem{definition}{Definition}

\newcommand{\showcomments}{true}
\newcommand{\gui}[1]%
{\ifthenelse{\equal{\showcomments}{true}}%
{{\color{magenta}{\small [\textbf{gui:} #1}]}}{\xspace}}%

\newcommand{\Hcal}{\mathcal{H}}
\newcommand{\Tcal}{\mathcal{T}}

\newcommand{\Rcal}{\mathcal{R}}
\newcommand{\Scal}{\mathcal{S}}
\newcommand{\Ocal}{\mathcal{O}}

\begin{document} 

\pagenumbering{roman}

\begin{titlepage}
\setlength{\topmargin}{1in}
\baselineskip=15.5pt \thispagestyle{empty}


\begin{center}
    {\fontsize{19}{24}\selectfont \bfseries \emph{To be or not to be}, but where?}
\end{center}

\vspace{0.1cm}
\enlargethispage{-2cm}
\renewcommand*{\thefootnote}{\fnsymbol{footnote}}

\begin{center}
    {\Large{Guilherme Franzmann}$^{1,2,3}$ \footnote{guilherme.franzmann@su.se}  }
\end{center}

\renewcommand*{\thefootnote}{\arabic{footnote}}
\setcounter{footnote}{1}

\begin{center}
\vskip8pt
   \noindent 
    \textsl{$^1$Nordita,
KTH Royal Institute of Technology and Stockholm University,
\\ Hannes Alfvéns v\"ag 12, SE-106 91 Stockholm, Sweden}

    \vskip8pt
   \noindent 
    \textsl{$^2$Department of Philosophy, Stockholm University, \\Universitetsvägen 10D, 114 18 Stockholm, Sweden}
    
     \vskip8pt
   \noindent 
    \textsl{$^3$Basic Research Community for Physics e.V., \\Veddeler Brückenstrasse 122, 20539 Hamburg, Germany}

\end{center}

\vspace{1cm}

\hrule
\vspace{0.3cm}

\noindent {\bf Abstract} 

The identification of physical subsystems in quantum mechanics as compared to classical mechanics poses significant conceptual challenges, especially in the context of quantum gravity. Traditional approaches associate quantum systems with classical ones localized in spacetime, using either Hilbert space factors for finite-dimensional systems or local operator algebras in algebraic quantum field theory. These methods ensure statistical independence for state preparations and measurements. However, covariant linearized quantum gravity disrupts this framework by preventing the formation of gauge-invariant local algebras, thereby undermining statistical independence. This presents a major obstacle for modeling early universe cosmology and gravity-induced entanglement experiments, and poses a significant roadblock toward a comprehensive theory of quantum gravity. A pivotal shift is proposed: the identification between classical and quantum systems should be \textit{dynamically} evolving rather than fixed, opening the possibility of a single-world unitary quantum mechanics. This perspective aligns with the broader aim of understanding how classical spatiotemporal existence emerges from quantum mechanics and connects the measurement problem with quantum gravity.
\vskip10pt
\hrule
\vskip10pt

\newpage
\setlength{\topmargin}{0in}
\tableofcontents

\end{titlepage}

\clearpage



\pagenumbering{arabic}
\setcounter{page}{1}

\newpage

\input{introduction}

\input{tale_emergence}

\input{ask_philosopher}

\input{status_of_art.tex}

\input{gravity.tex}

\input{why_care}

\input{inside}

\section{Acknowledgements}
\input{acknowledgements}

\appendix

\phantomsection
\addcontentsline{toc}{section}{References}

\let\oldbibliography\thebibliography
\renewcommand{\thebibliography}[1]{
  \oldbibliography{#1}
  \setlength{\parskip}{0pt}
  \setlength{\itemsep}{0pt} 
  \footnotesize 
}

\bibliographystyle{JHEP2}
\bibliography{references}

\end{document}

%% file: introduction.tex

\section{The mysterious box}\label{sec:mysterious}

One day, you arrive at your lab and find a mysterious box. The box is completely black except for a small display where it says ``{\fontfamily{qcr}\selectfont
ENERGY SPECTROMETER}". You try to open the box, but cannot quite figure out how. So, you decide to interact with the display and write down all the possible energy values of whatever is inside the box. As you near the end of this tedious task, all the years spent modeling physical systems give you a hunch that perhaps whatever is inside the box is a simple physical system.

You decide to model it by using a lattice of spin$-\frac{1}{2}$ sites, or \emph{qubits}. Among all the different terms you can add to the Hamiltonian of this system, you start by considering that each site couples to its neighbor's spin in a given direction with some coupling strength $J$ and that they might be under the influence of some external magnetic field $h$. Given all your experience, and the fact that you know how many energy levels there are\footnote{Knowing how many energy levels there are, you also know the minimum size of the Hilbert space needed to model the system, assuming the absence of energy degeneracies.}, you propose the following model\footnote{We suppress identity operators for simplicity unless needed to make a point.}: 
\begin{equation}
    H = J\sum_{I=1}^{3} \sigma_I^z \sigma_{I+1}^z + h\sum_{I=1}^4 \sigma_I^z \,. \label{eq:Ising_model1}
\end{equation}
You believe that there are 4 qubits, each coupled to its neighbor in the $z$-direction, with a magnetic field in the same direction. You compute the energy spectrum of this system, and, to your surprise, there is a pair $\{J,h\}$  for which the spectrum matches exactly the values displayed in the black box.

Excited, you leave and call a friend to see the black box and share your discovery. When you both return to the lab and you are about to show your model, your friend stops you, wanting to figure out the mystery themselves. After some time, they return, having found their model for what is inside the box, described by the following Hamiltonian:
\begin{equation}
       H = J\sum_{I=1}^{4} \mu^x_I +  h\sum_{I=1}^{3} \mu_I^z \mu_{I+1}^z - J\mu_4^x + h\mu_1^z\,. \label{eq:Ising_model2}
\end{equation}
It is also a model of spin sites! It looks quite similar to your own, except that the coupling between their spin sites is not the same as yours, but what you had as the magnetic field. And their magnetic field is along the $x$-direction, and not over all their spins... also, their first spin is under another magnetic field in the $z$-direction. How is that possible? 

You discuss what is going on, trying to understand how you both arrived at different models of what is inside the box that nonetheless matched exactly the energy spectrum. Until your friend has an idea: what if what you both called qubits are somehow related to each other? After some thought, you arrive at the following identification:
\begin{equation}\label{eq:relation_sites}
    \mu_I^z = \prod_{J\leq I}\sigma_J^x, \quad \mu_I^x = \sigma_I^z \sigma_{I+1}^z,\quad \mu_4^x = \sigma_4^z\,, 
\end{equation}
which, together with the swapping of the coupling strength and magnetic field, maps one model exactly onto the other one. 

At first, you are not very surprised, as it is quite common in physics to find different models to describe the same phenomenology. However, as you both ponder this map, you start to become suspicious of what is going on. For instance,  
\begin{equation}
    \mu^z_4 = \sigma_1^x \sigma_2^x \sigma_3^x \sigma_4^x \,,
\end{equation}
tells you that spin site $4$ in the $\mu$-model in the $z$-direction is somehow mapped onto the product of all the spin sites in the $x$-direction in the $\sigma$-model. Meanwhile, 
\begin{equation}
    \mu_1^x = \sigma_1^z \sigma_2^z\,,
\end{equation}
relates the $\mu$-sites in the $x$-direction to pairwise neighbors in the $\sigma$-model in the $z$-direction. The problem is that both you and your friend used the numbers ${1,2,3,4}$ to label spin \emph{sites}, which are localized in spacetime. However, the spin sites in each description are non-locally related. Each model assumes a different notion of the system, interactions, and spatial distribution of the physical subsystems. But surely, one model must be the \emph{real} one, or so you hope.

As you both start to lose patience and commit to discovering what the box hides, you decide to follow the example of particle physicists and grab a hammer to smash it open. Just as you are about to strike the first blow, the box disappears. Nevertheless, the question remains: How can you make sense of what was inside the box?

\section{How space hides in quantum mechanics}\label{sec:sapce_QM}

The \textit{gedanken} experiment above illustrates that quantum systems only described by their global Hilbert space, $\Hcal$, a state vector, $\ket{\Psi}$, and a Hamiltonian, $H$, might not fully characterize its classical properties, such as the notion of a ``site" -- a location in space \cite{Stoica:2024nhh}. After all, we only had access to the overall energy associated with each of the global states of the system\footnote{Once again, we considered a simplified scenario where each energy value displayed in the spectrometer is associated with a non-degenerate energy eigenstate, thus also fixing the size of the Hilbert space. All the eigenvalues together can be used to reconstruct the Hamiltonian matrix in its diagonal form.}. This is puzzling because the situation is typically reversed: we impart classical notions of location into our quantum mechanical descriptions. For example, for every spin site located at position $I$, we assign a $\mathbb{C}^2$-Hilbert space factor, such that the global Hilbert space is
\begin{equation}
    \Hcal = \bigotimes_{I=1} \mathbb{C}^2\,.
\end{equation}
This is just a simple example of a general attitude\footnote{We can speculate whether this attitude is intrinsically derived from the quantization program, where classical observables are promoted to quantum operators, and among them, position and momentum take a special role. Had the program been \emph{classicalization} instead, then the emphasis could have been rather different.}: we typically assign a Hilbert space $\Hcal_{\rm system}$ to each classical subsystem in a given spatial (or more generally, spacetime) region\footnote{As usual, this story is fairly more complicated, in particular in the context of infinite-dimensional systems and constrained theories, where only the notion of localizability for the observables is guaranteed \cite{Fewster:2023cfq}. We will come back to that in Sec. \ref{sec:status_art} and \ref{sec:gravity}.} $U$. Thus, some notion of space, or at very least, some notion of location in space, seems to leak into our quantum-mechanical descriptions. Let's make that explicit. 

Back at your lab, you consider an extended quantum mechanical system divided into two parts, $A$ and $B$, with a global Hilbert space $\Hcal$. They can be interacting or not, very much like the spin sites from above. Each part is surrounded by measuring devices, defining the sets of controlled observables $\{\Ocal^A\}$ and $\{\Ocal^B\}$, which belong to the algebra of operators on $\Hcal$, $\mathcal{L}(\Hcal)$, while a clock fixed against the wall sets a reference frame for the entire lab. Thus, at any fixed time in the lab, these parts are spacelike separated, and you expect measurements on $A$ to not affect measurements on $B$, at least within the time interval light takes to travel between the systems. Essentially, you aim to ensure that the statistics of these parts are independent, implementing an operational notion of macrocausality. For example, the statistics associated with any observable in part $A$ are obtained by $\tr [\rho \Ocal^A ]$, where $\rho$ is the state of the system and belongs to the state space $\Scal(\mathcal{L}(\Hcal))$. Now consider two measurements, one in each part of the system, defining two spacetime events that are spacelike separated. Each measurement updates the state of the system, but their local (corresponding to one of the subsystems) statistics should be the same whether we consider the original state, $\rho$, or the updated state corresponding to one of the measurements being performed, $\rho'$, since they were performed at spacelike separations. Therefore, we demand $\tr[\rho\Ocal^A ]=\tr[\rho' \Ocal^A ]$, and similarly for $B$. It turns out that this is possible if, and only if, the set of observables for each part commute \cite{Malament1996-MALIDO,EmerGe_proj_2}, i.e. $[\Ocal^A,\Ocal^B]=0$. This argument can be generalized to subalgebras associated with arbitrary causally disconnected spacetime regions, where then goes by the name \emph{microcausality}, or Einstein separability. We will discuss this in Sec. \ref{sec:status_art}. 

Thus, we have two commuting sets of observables. If we also require that they together span all the possible operators associated with the system\footnote{Remember that any operator can be decomposed into a complex sum of two self-adjoint operators, i.e., observables.}, i.e., $\Ocal^A \otimes \Ocal^B \cong \mathcal{L}(\Hcal)$, then it follows \cite{Zanardi:2004zz} that the Hilbert space of the quantum system can be factorized, $\Hcal = \Hcal_A \otimes \Hcal_B$. Thus, our local assignment of Hilbert space factors to parts of a bigger system is born out of macrocausality, operationally implemented in a way that leads to microcausality, and finally leading to the factorization of the Hilbert space\footnote{Of course, most of the time, we do not think in these terms as we consider non-relativistic quantum mechanics. Then, the factorization of the Hilbert space is simply meant to characterize the independence between quantum subsystems, regardless of a consistent spacetime embedding where state preparations and measurements are \textit{causally} independent.}. 

Now that the physical intuition is properly encoded in the math, let's be more precise and general. Following \cite{Zanardi:2004zz}, we consider a finite-dimensional Hilbert space $\Hcal$ and a collection of subalgebras of observables corresponding to each subsystem, $\mathcal{A}_I$, satisfying the following properties: 

\begin{itemize}
    \item \textit{Local acessibility}: Each $\mathcal{A}_I$ corresponds to a set of controllable observables;
    \item \textit{Subsystem independence}: $[\Ocal^I,\Ocal^J]=0$ for $I\neq J$, where $\mathcal{O}_I \in \mathcal{A}_I$;
    \item \textit{Completeness}: $\bigvee_I \mathcal{A}^I \cong \mathcal{L}(\Hcal)$, there is a minimal inclusion of the subalgebras that is isomorphic to the whole algebra of operators.
\end{itemize}
Once these properties are in place, then the set of subalgebras induces a tensor product structure (TPS) \cite{Cotler:2017abq}: 

\begin{definition}[TPS] \label{def:TPS}: A TPS $\Tcal$ of Hilbert space $\Hcal$ is an equivalence class of isomorphisms $T:\Hcal \rightarrow \otimes_I \Hcal_I,$ where $T_1 \sim T_2$ whenever $T_1 T_2^{-1}$ can be written as a product of local unitaries $\otimes_I U_I$ and permutations of subsystems.
\end{definition}

By choosing a TPS, we determine how the Hilbert space of the entire quantum mechanical system (like the box in the example above) will be partitioned. However, now we can understand that this assignment is intrinsically related to our classical notion of spacetime. On the other hand, if nature is fundamentally quantum mechanical, it might make more sense to start with the global Hilbert space and then find an appropriate decomposition\footnote{There are recent works in this direction, for instance \cite{Carroll:2020gme,Adil:2024wok,Cotler:2017abq}, while \cite{Piazza:2005wm} poses for the first time the question of which TPS corresponds to the partition of the Hilbert space into ``regions of space''.} - particularly one that aligns with our classical observables, such as the notion of \emph{site 1 is here, site 2 is there}. The question, however, is whether there are multiple classical possibilities associated with the same quantum system, as seen in the different \emph{site} attributions in our gedanken experiment above, and vice-versa.

Once a TPS is introduced, we examine the 3-tuple $(\Hcal, H, \ket{\Psi})$ through the composition of each factor $\Hcal_I$ (or local subalgebras, $\mathcal{A}_I$). This is exactly what happened in our box experiment, where you and your friend tacitly assumed different TPSs. Each $\Tcal$ considered is not related by local unitaries or permutations of subsystems; instead, the set of variables is non-locally related as shown in \eqref{eq:relation_sites}, amounting to a global unitary transformation. Thus, the TPSs are not equivalent. That can also be checked by fixing the factorization of the Hilbert space, $\bigotimes_I \Hcal_I$ and showing that for each TPSs $\Tcal_1$ and $\Tcal_2$, the operators $T_1 HT_1^{-1}$ and $T_2 H T_2^{-1}$ are not the same up to conjugation by local unitaries, permutations of subsystems, and transposition. Therefore, a change of TPS has an equivalent perspective where the TPS is fixed but the Hamiltonian is unitarily transformed, keeping its spectrum \cite{Cotler:2017abq}. 

Now, although the TPSs implicitly assumed for each description of the box are not equivalent, the associated Hamiltonians present the same level of locality, being the sum of $1$- and $2$-point terms. Both Hamiltonians are said to be $2$-local. This is a specific case of a duality \cite{Cotler:2017abq}: 

\begin{definition}[Duality] \label{def:duality}
Two TPSs are dual if they are not equivalent and their associated Hamiltonians, which have the same spectra, are $k$-local. 
\end{definition}

You might expect that the issue of identifying the appropriate classical limit, understood as the assignment of Hilbert space factors to spatiotemporally localized systems, arises only when dual descriptions exist. After all, if the box cannot be opened (i.e., we cannot extend its Hamiltonian \cite{Huggett:2020cst}), we have incompatible classical descriptions of what is in the box. However, the problem runs deeper. Even without dualities, a generic Hilbert space with a generic Hamiltonian will have multiple tensor factorizations, only constrained by the dimension of the Hilbert space. And without further specification, all these decompositions are equally reasonable representations of the quantum-mechanical system \cite{Stoica:2024nhh,DiBiagioFranzmannSoulas}. 

This quantum democracy has led to recent efforts to identify criteria for singling out a preferred factorization of the Hilbert space \cite{Carroll:2020gme,Adil:2024wok}. In particular, \cite{Cotler:2017abq} suggests a criterion based on the fact that for a TPS with $N$ subsystems, whenever $N \gg k$, the set of $k$-local Hamiltonians has measure zero. If one can then prove that dualities are not present, this would imply an almost unique TPS for a $k$-local Hamiltonian. Thus, by considering $k$-locality of the Hamiltonian as an additional structure to $(\Hcal, H, \ket{\Psi})$, a preferred factorization is selected. This also determines a preferred way to map quantum subsystems, as factors in Hilbert space, into classical ones, as systems localized in spacetime\footnote{For a critical view of these results, see \cite{stoica2023no,Stoica:2024nhh,DiBiagioFranzmannSoulas}}.

The keyword here is \emph{map}. As we have discussed, there are ways to connect aspects of the classical world to its underlying quantum structure. For example, we have established the connection between operational macrocausality and microcausality, and discussed how classical systems are associated with algebraic structures in quantum mechanics. However, we have also encountered dualities that hinder this direct connection. To further explore how the classical world relates to the quantum one, we will consider their connection through the lens of emergence.

%% file: tale_emergence.tex
\section{A Tale of Emergence}\label{sec:emergence}

In this section, we will discuss the concept of emergence, start to clarify the distinction between classical and quantum systems, and establish the foundations to better understand our mysterious box. 

We start by providing our working definition of emergence. Following \cite{Butterfield_2010,Butterfield_2011,Oriti:2018dsg}, we consider: 

\begin{definition}[Emergence] \label{def:emergence}
Emergence is understood as the appearance, in a certain description of a physical system, of properties that are novel with respect to a different (more ``fundamental") description of the same system, and robust, thus stable enough to represent a characterization of the new description and to form part of new predictions stemming from it. Emergence, in this understanding, usually requires the use of some limiting procedure and of a number of (possibly) drastic approximations, to allow the novel properties to become visible in the new description. 
\end{definition}

Key to this definition are the terms \emph{novel}, \emph{robust}, and \emph{new predictions}\footnote{On the other hand, one could think that we want to avoid such descriptions by being vague, and seek out definitions of emergence that are strictly mathematical. A recent work in that direction is \cite{Carroll:2024nxc}. We plan to consider this approach in future works.}. When we have a description that is novel and robust, in the sense that it stands on its own and provides a way to predict the behavior of physical systems, we typically call it a physical theory. Of course, this also underlies what we mean by a more fundamental description. Therefore, perhaps we should stop here and scrutinize the structure of what we mean by physical theories to make the different flavors of emergence more explicit. 

\subsection{The Structure of Physical Theories}\label{sec:physical_theories}

Most physical theories can be broken into four different categories. Briefly,  
\begin{itemize}
\item \textbf{Underlying state space} ($\Omega$): This represents the space of all possible states of a physical system. For example, in Lagrangian mechanics, this is the configuration space, while in Hamiltonian mechanics, it is the phase space. In quantum mechanics, we typically consider the Hilbert space\footnote{See footnote \ref{ft:states}, though.};
\item \textbf{Degrees of freedom} ($\Phi$): For the same type of underlying space, there can be different degrees of freedom. For instance, in phase space, one can describe a particle or a field. In quantum mechanics, we typically consider the wavefunction (state vector) or, more generally, the density matrix;
\item \textbf{Observables} ($\mathcal{O}$): These are the quantities that can be measured, typically represented as maps from the state space to the set of real numbers\footnote{This is the naive definition. The more general definition flips the role of states and observables, and it is the state that is a linear functional of the observables into the real numbers, as introduced later in the paper.};
\begin{itemize}
\item In classical mechanics, observables are typically functions  $f: \Omega \rightarrow \mathbb{R}$ , where  $\Omega$  is the phase space, for example.
\item In quantum mechanics, observables are self-adjoint (Hermitian) operators  $\Ocal$  on a Hilbert space  $\mathcal{H}$, ensuring real eigenvalues that are associated with measured values. 
\end{itemize}
\item \textbf{Dynamical equation} ($U$): Once a state is known for a given degree of freedom, the dynamical equation determines the evolution of the state, defining a trajectory in the state space. In classical mechanics, we have Newton's, Euler-Lagrange, or Hamiltonian equations\footnote{In this section, we will assume classical mechanics is governed by Newton’s equations, as this is sufficient to make our point.}, while in quantum mechanics, we have the Schrödinger or Heisenberg equation.
\end{itemize}
In the language of \cite{deHaroButterfield}, these elements are reduced to a triple $T=(S,Q,D)$: the state-space, the quantities, and the dynamics. The reason why we say \emph{most} physical theories is that some do not fit this decomposition: e.g., in thermodynamics, the second law does not take the form of a dynamical equation. Nonetheless, to convey our points in the next sections, we will ignore these subtleties. 

We will also consider another simplification: instead of discussing the state space, we will focus on the notion of \emph{stage space ($\mathcal{S}$)}, which is the space implicitly or explicitly associated with some notion of existence, serving as the background against which a theory is defined. Thus, to discuss emergence, we will rely on the scheme defined by the quartet $(\mathcal{S},\Phi, \mathcal{O},U)$. 

Now that we have at hand the definition of emergence and the underlying structure of physical theories, we can establish different levels of emergence by considering some examples.  

\subsection{Emergence - Level 0: Classical Mechanics and the Center of Mass}\label{sec:emergence_lvl_0}

Our first example involves the center of mass of a system. The center of mass is an important concept in classical mechanics because it simplifies the analysis of many-body systems. By reducing the complexity of the system, we can describe the motion of the entire system using a single point that represents the average position of all the particles weighted by their masses. This simplification is useful in various applications, such as understanding the motion of celestial bodies, analyzing the stability of structures, and simplifying collision problems. 

We begin by defining the center-of-mass particle's position as a map $X_{\rm CM}$,
\begin{align}
    X_{\rm CM}: 
    \left\{ 
    \begin{aligned}
    &\mathds{R}^{3N} \longrightarrow \mathds{R}^{3} \nonumber \\
    &\vec{x}_{\rm CM} \equiv  \frac{1}{M} \sum_{I=1}^N m_{\rm I}  \vec{x}_{\rm I}\,, \quad \quad M= \sum^N_{I=1} m_{\rm I}\,,
    \end{aligned}
    \right.
\end{align}
which maps position vectors in spacetime to a single position vector in spacetime\footnote{If we wanted to focus on state space instead of spacetime, we could say that this map collapses the $N$-particle configuration space into a 1-particle configuration space.}. This map is surjective but not injective, so it is not invertible. In other words, it is a \emph{projection map}. This means that if we only had access to the center of mass data of a system, there would be many inequivalent possible lower-level arrangements of it that would correspond to that information. 

Thus, we can consider the center of mass as a new description of the system, even though the type of degree of freedom remains the same. The system is now described by the position and velocity of the center-of-mass particle. Since the type of degree of freedom remains the same in the new description, the type of observables also remains the same, namely differentiable functions, ${f_a}$, that map points in phase space to real numbers (or probability distributions, more generally). Thus, we can also think of the center-of-mass map as inducing a projection map in the space of observables, which changes the codomain of the observables from $\mathds{R}^{3N}\times\mathds{R}^{3N}$ to $\mathds{R}^3\times\mathds{R}^3$. However, the type of observables does not change, meaning we can still ask the same questions about the center of mass as we could about any of the particles of the system before the mapping. The same applies to the equation of motion, which remains Newton’s second law for a particle with total mass $M$, resulting from the linearity of the equation.

We summarize all these points in Table \ref{tab:CM} and take it to represent a mild form of emergence, referred to \emph{Emergence - Level 0}. 

\begin{table}
\captionsetup{font=small}
    \centering
\begin{tabularx}{1 \textwidth} { 
   >{\centering\arraybackslash}X
   >{\centering\arraybackslash}c
   >{\centering\arraybackslash}X
   >{\centering\arraybackslash}c
   >{\centering\arraybackslash}X }
 \toprule
& & \small{\textbf{Fundamental Description}} &  & \small{\textbf{Emergent Description}} \\ \midrule 
 \small{\textbf{Stage Space}} & & $\mathcal{M}$   &    &  $\mathcal{M}$  \\ \\
    \small{\textbf{Degrees of Freedom}} & & $10^{51} \text{ of } (\vec{x}, \vec{v})'s$  &   & $(\vec{x}_{\rm CM}, \vec{v}_{\rm CM})$ \\ \\
     \small{\textbf{Observables }}& & $f_a(\vec{x}_{\rm I}, \vec{v}_{\rm I})$   &   &  $f_a(\vec{x}_{\rm CM}, \vec{v}_{\rm CM})$ \\ \\
    \small{\textbf{Dynamical Equations}} & & $\vec{F}_I(\vec{x}_{\rm I}) = m_{\rm I} \frac{d\vec{v}_{\rm I}}{dt}$  &  & $\vec{F}(\vec{x}_{\rm CM}) = M \frac{d\vec{v}_{\rm CM}}{dt}$\\
\bottomrule
\end{tabularx}
    \caption{ Summary of the fundamental and emergent descriptions for the center-of-mass of an object in classical mechanics. Fundamentally, there is a large number of degrees of freedom for individual particles - for instance, Earth alone has approximately $10^{51}$ particles. The emergent description retains the same type of observables and equations of motion but reduces the complexity by aggregating the system into a single center-of-mass particle. The stage space remains the same.}
    \label{tab:CM}
\end{table}

\subsubsection{Does the center of mass exist?}

As we aim to build our intuition around ontology as emergence unfolds, we can begin by asking some fundamental questions. There is plenty of literature around this topic \cite[e.g. and references therein]{wallacerealpatterns}, but we just want to highlight a few things to slowly build our intuition for later on.

First, in some situations, the center of mass defines a point that lies within the composite object it represents, making it materially instantiated. In such cases, there is literally a material point with its own position and velocity vector that follows the center-of-mass trajectory. However, this is not always the case. Often, the center of mass does not correspond to any actual material point within the defining object. Nonetheless, from a substantivalist perspective, all points in space exist in an absolute sense in any case, so the center of mass trajectory simply labels some of these points.

Rather than focusing on material instantiation or substantivalism to argue for the existence of the center of mass, we can adopt the perspective supported by our definition of emergence. As long as new entities are novel and stable enough to yield reliable predictions, as is the case for the center of mass, we can attribute emergent ontological status to them, similar to Wallace's implementation \cite{wallacerealpatterns} of Dennett's \emph{real patterns} \cite{Dennett1991-DENRP} in physics. 

\subsection{Emergence - Level 1: Classical Mechanics and Coarse Graining}\label{sec:emergence_lvl_1}

Our next example is a well-established one: how thermodynamics emerges from statistical mechanics. We will simplify the discussion by considering only an ideal gas in equilibrium, focusing on how the different elements of the physical structure change from one theory to the other. 

Thermodynamics also applies to systems of many particles, but instead of averaging over the positions of the particles, we average over the square of their velocities. We define the mean map $M$ as follows:
\begin{align}
   M : 
   \left\{
   \begin{aligned}
       &\mathds{R}^{3N} \longrightarrow \mathds{R} \\
   & \langle v^2 \rangle \equiv \frac{1}{N} \sum_{I=1}^N v_I^2\,,
   \end{aligned}
   \right.
\end{align}
from which energy, temperature, and pressure for an ideal gas are defined as,
\begin{equation}
    E= \frac{3}{2}k_B \langle v^2 \rangle\,,\quad \quad T = \frac{m}{3k_B} \langle v^2 \rangle\,, \quad \quad P = \frac{Nm}{3V} \langle v^2 \rangle \,.
\end{equation}
Here,  $V$ is the volume of the container where the particles can move, and $m$ is their mass. 

Together, the equations above give rise to the ideal gas law, $PV=k_B n T$, where $n$ is the number of moles, which is a significant departure from the underlying Newton's equation governing each of the particles in the gas. In the equilibrium case, the situation is quite drastic as all the degrees of freedom are reduced to three, but in reality, two, since the ideal gas law is a constraint equation that establishes the temperature and pressure of a uniform gas in a box of volume $V$. Thus, all the physical observables associated with the box are now functions of these macroscopic quantities, $F_a(T,V,P)$, which are quite different from the underlying particle ontology, since we now talk about entropy, heat capacities, etc. 

Despite the change in the type of degrees of freedom, equations, and observables, the stage space remains physical space. To make this more explicit, we can relax the condition of global thermal equilibrium and consider only local thermal equilibrium. In this case,  the ideal gas law holds locally, and the temperature and pressure become scalar fields that vary from point to point in space. In this scenario, there are also dynamical equations, known as general transport equations, that explicitly incorporate time. 

We summarize this example in Table \ref{tab:idealgas} to illustrate what we call \emph{Emergence - Level 1}. 

\begin{table}
\captionsetup{font=small}
    \centering
\begin{tabularx}{1 \textwidth} { 
   >{\centering\arraybackslash}X
   >{\centering\arraybackslash}c
   >{\centering\arraybackslash}X
   >{\centering\arraybackslash}c
   >{\centering\arraybackslash}X }
 \toprule
  & & \small{\textbf{Fundamental Description}} &  & \small{\textbf{Emergent Description}} \\ \midrule 
 \small{\textbf{Stage Space}} & & $\mathcal{M}$   &    &  $\mathcal{M}$  \\ \\
    \small{\textbf{Degrees of Freedom}} & & $n\times10^{23} \text{ of } (\vec{x}, \vec{v})'s$  &   & $(T,V,P)$ \\ \\
     \small{\textbf{Observables}} & & $f_a(\vec{x}_{\rm I}, \vec{v}_{\rm I})$   &   &  $F_a(T,V,P)$ \\ \\
    \small{\textbf{Dynamical Equations}} & & $\vec{F}_I(\vec{x}_{\rm I}) = m_{\rm I} \frac{d\vec{v}_{\rm I}}{dt}$  &  & $PV=k_{\rm B} n T$\\
\bottomrule
\end{tabularx}
    \caption{Summary of the fundamental and emergent descriptions in the context of equilibrium thermodynamics. Once again, fundamentally there is a large number of degrees of freedom for individual particles (of order $n$ times Avogadro's number in a room), while the emergent description simplifies the system by focusing on macroscopic thermodynamic variables like temperature, volume, and pressure. The emergent description retains the same stage space but significantly reduces the complexity by aggregating the behavior of numerous particles into a few macroscopic observables and a single equation.}
    \label{tab:idealgas}
\end{table}

\subsubsection{Does temperature exist?}

We can now pose the same type of question we asked about the center of mass, but this time for the temperature field. If the ontological status of the center of mass was unclear but did not prevent us from using emergence to assign it a sense of reality at the emergent level, the same applies to the temperature field.

In fact, the more \emph{drastic approximations} we impose to transition from the base theory to the emergent description, the more inclined we might be to establish the independent existence of entities at the emergent level, as long as they help provide reliable predictions. This is certainly the case for the temperature field, as it is a key variable in equations of state, such as the ideal gas law, and in determining phase transitions and thermal properties. Moreover, temperature is a directly measurable quantity; it can be empirically determined using thermometers and other temperature-sensitive devices. These measurements consistently align with theoretical predictions, reinforcing the reality of temperature as an emergent property. Thus, while temperature is not a fundamental property at the microscopic level, its emergent ontological status is well-supported by its predictive power and experimental verification.

\subsection{Emergence - Level 2: Quantum Mechanics and Classical Mechanics}

Just a year after Boltzmann equations were derived to describe the statistical behavior of out-of-equilibrium thermodynamical systems, Riemann \cite{riemann} was already anticipating what would be required for the next level of emergence:

\begin{quote}
    The questions about the infinitely great are for the interpretation of nature useless questions. But this is not the case with the questions about the infinitely small. It is upon the exactness with which we follow phenomena into the infinitely small that our knowledge of their causal relations essentially depends. \\ 
    (...)\\
    Now it seems that the empirical notions on which the metrical determinations of space are founded, the notion of a solid body and of a ray of light, cease to be valid for the infinitely small. We are therefore quite at liberty to suppose that the metric relations of space in the infinitely small do not conform to the hypotheses of geometry, and we ought in fact to suppose it, if we can thereby obtain a simpler explanation of phenomena. (Riemann, 1873) 
\end{quote}
And about three decades later, quantum mechanics was developed, and the stage space itself would no longer be based on spacetime, but rather on quantum space: the Hilbert space. 

Quantum mechanics also follows the structure of physical theories discussed above. Quantum mechanical degrees of freedom are typically represented as vectors in Hilbert space\footnote{In this section, this simplification suffices for the points we want to make.}, evolving according to the Schrödinger equation. Physical observables are represented by self-adjoint operators instead of smooth measurable functions. However, at face value, the theory introduces other structures, such as the Born rule and the (real or apparent) collapse of the quantum state: the Schrödinger equation is linear, it allows for superpositions of different states, but it seems that measurements unavoidably reduce the superimposed quantum state to a single one. In other words, measurements seemingly give out single outcomes. This is also known as the \emph{measurement problem}\footnote{Given the extensive literature, among personal favorites are \cite{brukner2015quantum,Mermin:1996mr,wallace2021philosophy,landsman2017foundations}} and will be further discussed in Sec. \ref{sec:O-W_UQM}. 

Now, it is generally believed that the world is fundamentally quantum mechanical. After all, it rules the small, microscopic world, and the macroscopic world is composed of these small entities, so it seems that classical mechanics has lost its status as a base theory, and should be seen as the emergent description of the macroscopic world instead. As this time around the state space also changes, we call this \emph{Emergence - Level 2}. We summarize the example of the classical emergent description from quantum mechanics in Table \ref{tab:class_quant}. 

\begin{table}
\captionsetup{font=small}
    \centering
\begin{tabularx}{\textwidth} { 
   >{\centering\arraybackslash}X
   >{\centering\arraybackslash}c
   >{\centering\arraybackslash}X
   >{\centering\arraybackslash}c
   >{\centering\arraybackslash}X }
 \toprule
 & & \small{\textbf{Fundamental Description}} &  & \small{\textbf{Emergent Description}} \\ \midrule 
 \small{\textbf{Stage Space}} & & $\mathcal{H}$   &    &  $\mathcal{M}$  \\ \\
   \small{\textbf{ Degrees of Freedom}} & & Q-states $|\psi\rangle $  &   & $(\vec{x}_{\rm I}, \vec{v}_{\rm I})$ \\ \\
    \small{\textbf{ Observables}} & & $\hat{\mathcal{O}}_A^\dagger = \hat{\mathcal{O}}_A$   &   &  $f_a(\vec{x}_{\rm I}, \vec{v}_{\rm I})$ \\ \\
    \small{\textbf{Dynamical Equations}} & & $i\hbar \frac{d}{dt} |\psi\rangle = \hat{H} |\psi \rangle$   &  & $\vec{F}_I(\vec{x}_{\rm I}) = m_{\rm I} \frac{d\vec{v}_{\rm I}}{dt}$\\
\bottomrule
\end{tabularx}
    \caption{Summary of the fundamental and emergent descriptions in the context of quantum and classical mechanics. At the quantum level, the stage space is represented by the Hilbert space, and self-adjoint operators act as observables on quantum states that follow the Schrödinger equation. Meanwhile, the classical stage space is spacetime, where we define positions and velocities and measurable functions of them as observables, and the dynamical equation is Newton's equation.}
    \label{tab:class_quant}
\end{table}

\section{To be or not to be, but where?}

If we had doubts about the existence of emergent entities in the previous examples of lower levels of emergence, it is now time to question or reaffirm the existence of classical degrees of freedom, and even their stage space, through the lens of emergence. Unlike the previous examples, degrees of freedom and observables associated with space and time, and space and time themselves, are deeply rooted in our physical intuition about the world. 

This is evident from our language. In fact, ontology, understood as the study of what there are and their relations \cite{sep-logic-ontology}, relies on the verb \emph{to be} for its definition. This verb is inherently intransitive and can stand alone to define existence. However, it reflects the physical intuition that entities exist in space and time, as no other type of stage space was considered until the advent of quantum mechanics, at least in physics\footnote{In the philosophy of mind, non-physicalist philosophers have introduced mental states that do not have the material world as their stage space. The ontological challenges of reconciling the classical and quantum stage spaces can draw parallels from that literature, as explored in \cite{LeBihan2018-LEBPMB, lebihanhard}, where the \emph{hard problem of spacetime} is discussed.}. Thus, we are now faced with two possibilities: i) to move from a classical ontology (where the verb \emph{to be} is defined in spacetime) to a quantum ontology (where the verb \emph{to be} is defined in Hilbert space), or ii) to allow for a multi-level ontology \cite{Oriti:2018dsg}, reassured by the notion of classical real patterns emerging from the underlying quantum world \cite{wallacerealpatterns}. 

We could summarize the second perspective by promoting the verb ``to be" from an intransitive verb to a mandatory prepositional verb, where things \textit{are} \emph{in spacetime} or \emph{in Hilbert space}. By establishing this nuance in language, it becomes clearer what work needs to be done to connect the entities defined in each space. In particular, as seen in the previous examples of lower levels of emergence, lower-level entities can typically be grouped in many different ways to give rise to entities in the emergent description (e.g., many configurations of masses can give rise to the same center of mass position). Thus, the typical identification of classical systems with their quantum counterparts seems to be a limitation of the quantization program, which starts from classical systems and attempts to derive their quantum description, instead of starting from the base level where quantum systems are defined and allowing a considerable amount of ambiguity in how their classical counterparts are obtained in the emergent level. It is this latter program, motivated by insights from quantum gravity, that we will pursue in Sec. \ref{sec:O-W_UQM}.

%% file: ask_philosopher.tex
\subsection{Where does the ontology reside?}

Returning to our initial example, we find ourselves in a conundrum: is the system inside the box a spin chain with interacting sites represented by $\sigma_I$ or $\mu_I$, or perhaps something else entirely? We might hope to break the box and directly observe its contents, thus extending the initial Hamiltonian and coupling the system to something else. However, if the box cannot be broken—similar to the lone harmonic oscillator and string theory discussed in \cite{Huggett:2020cst}—we face a choice: where should the ontology reside?

Before exploring the options, it is important to emphasize that we have identified two different notions of systems. Classical systems occupy regions of spacetime and are described by degrees of freedom in spacetime, while quantum systems are described by degrees of freedom in Hilbert space. While classical systems typically interact locally in spacetime, meaning their interactions are point-like in the Hamiltonian (the systems need to make contact), quantum systems interact locally when the interaction terms in the Hamiltonian can be factorized as the product of operators acting independently on each Hilbert space factor associated with each subsystem (for a review, see \cite{EmerGe_proj_2}). Thus, both the notion of subsystems and the notion of locality in the Hamiltonian are fundamentally different\footnote{The mismatch between quantum-mechanical locality and spatiotemporal locality has been discussed in the context of loop quantum gravity, referred to as disordered locality \cite{Markopoulou:2007ha}, and even considered as a possible explanation for dark energy \cite{Prescod-Weinstein:2009bqa}.}.

\subsection{What are the possibilities?}\label{sec:ontologies}

As ontology can be quickly summarized as the study of what there is, it is crucial to clarify what is meant by \emph{what} and \emph{is}. As we discussed above, we now see that \emph{what} can refer to two very different types of things: classical or quantum systems. Similarly, \emph{being} can be defined in different senses, either in terms of the classical or quantum stages. Moreover, the discussion also expands to the possible maps that could be established between these different modes of existence, which unavoidably leads to the discussion about the quantum-classical limit.

The project of ontology extends beyond simply defining what exists. Once we establish (or rather, posit) what things exist, the project then explores the most general features and relations among these entities. Therefore, our discussion about locality, and more broadly physical interactions, is also directly related to ontological matters. 

Another important aspect of this discussion is the expected plurality of entities. If we argue for the universe as a whole, either classically or quantum-mechanically, we talk about \emph{things} in the singular and focus on the \textit{holistic} aspects of either spacetime or Hilbert space, their global state, and Hamiltonian as what ultimately exists. The partition of these spaces into what we typically see as classical and quantum subsystems is viewed as emergent, despite the persistence of \emph{real patterns} associated with emergent entities \cite{Dennett1991-DENRP, wallacerealpatterns}. From this perspective, the second part of the ontological project becomes superfluous, as there is no meaningful notion of \emph{relations} for a set composed of a single element, except the trivial comparison with the empty set\footnote{An instance of this comparison in physics can lead to questions such as \emph{``Why is there something rather than nothing?''} \cite{Carroll:2018nul}.}. A recent defense of this position can be found in \cite{Carroll:2021aiq}. This holistic approach will be contrasted with the \emph{atomic} one, where there exists more than one single irreducible entity and relationships among them.

Thus, we end with four different possibilities: \{holism, atomism\} $\times$ \{classical, quantum\}. We have identified at least one relevant category to guide the atomistic approach, which is locality (see Table \ref{tab:ontology}). There is extensive literature discussing all these different possibilities, so we will not attempt this here, but just focus on crucial aspects involving these different ontologies that will set the stage for the rest of the paper.

\vspace{0.5cm}
\begin{table}[h]
\captionsetup{font=small}
\renewcommand{\arraystretch}{1.5} 
\begin{center}
\begin{tabular}{l@{\hspace{1em}}@{\hspace{1em}}c@{\hspace{1em}}@{\hspace{1em}}c@{\hspace{1em}}} 

\toprule
 & \small{\textbf{Quantum}} & \small{\textbf{Classical}} \\ 
\hline 
\small{\textbf{Holism}} & Holistic Quantum  Ontology &  Holistic Classical  Ontology\\
\hline 
\small{\textbf{Atomism}} & Atomistic$^{(k)}$ Quantum  Ontology & Atomistic Classical  Ontology \\ 
\bottomrule

\end{tabular}
\end{center}
\caption{Here we show the different possibilities where the ontology of the world can be placed. The index $(k)$ represents the degree of quantum-mechanical locality. Note that any degree of spatiotemporal nonlocality is seen as action-at-a-distance, while quantum-mechanical $k$-locality simply means that the interactions in the Hamiltonian involve terms relating up to $k$-subsystems, as discussed in Sec. \ref{sec:sapce_QM}.} \label{tab:ontology}
\end{table}

\subsection{Classical Ontology}

Why should we even discuss classical ontology? After all, isn't the world fundamentally quantum mechanical? This is precisely why we start with simpler questions about the ontological status of the center of mass and temperature. Similarly, we could invoke emergence to justify why classical mechanics should be granted ontological status, following the reasoning of \cite{wallacerealpatterns}. However, as we will discuss further in Sec. \ref{sec:status_art} and Sec. \ref{sec:O-W_UQM}, quantum mechanics, as it stands, appears unavoidably anchored in the classical world.

Moreover, we seem to exist in spacetime, so it is natural to align our ontological intuitions with this crucial aspect of our pursuit of increasingly better descriptions of the world. Ontology, after all, is meant to connect with our intuitive understanding of the world and it is reasonable to assign scientific reality (in the sense of scientific realism \cite{sep-scientific-realism}) to entities that take on clearly defined spatiotemporal regions. Let's consider each option carefully. 

\paragraph{Classical Holistic Ontology}
views the classical world as one. This perspective contrasts with the atomistic view, which holds that the world consists of independent parts. In classical physics, the theory that allows us to consider the world and its overall evolution is general relativity. In this framework, we can write equations, such as the Friedmann equations, that are meant to describe the entire Universe. It is by attributing reality to the solutions of these equations that we are led into discussions about \emph{eternalism} and the block universe \cite{carroll2011eternity}, as well as the cosmological multiverse \cite{diBiagioFranzmann}, for example. 

From an epistemological standpoint, classical holistic ontology challenges reductionism, which seeks to explain complex systems purely by understanding their individual parts \cite{Anderson:1972pca,wallacerealpatterns}. The properties of a whole system often emerge from the collective dynamics of its parts and cannot be predicted in practice solely from the behavior of individual components. Examples include the type of mechanical and thermodynamical emergent behaviors described in the last section. 

To some extent, this perspective was fully justified in Newtonian mechanics, where anything was non-locally connected to everything else. After special relativity, we learn that events can be spacelike separated, but as long as the background spacetime was Minkowsky, any two spacelike-separated events would always eventually find a common past. It is really with general relativity that this holistic ontology could face some challenges, as the very solutions of the theory that give rise to the cosmological multiverse describe distinct spacetime regions that are never in causal contact. This defines \emph{independent parts} of the Universe from a global perspective. Let's use this as a cue to move on to the next possibility. 

\paragraph{Atomistic Classical Ontology}

posits that the world is composed of distinct, independent entities that interact in spacetime. This perspective aligns with the traditional view in classical physics, where systems are considered as collections of discrete parts, each occupying a specific region in spacetime and interacting either locally or non-locally with other systems. Typically, we pursue classical theories that describe phenomena through localized interactions between particles or fields, emphasizing the independence of parts when regions are spacelike separated.

As discussed above, in general relativity there are solutions that establish clear global distinctions between different patches of spacetime. However, we often assume this to be possible, at least approximately, on a local scale. One can imagine a thick Cauchy slice of the entire spacetime and then consider any two spacelike-separated regions within this slice as independent. However, it is important to note that while events within these regions in a thick Cauchy slice cannot causally interfere with each other, the content of these regions might still be causally related through their past lightcone overlap\footnote{Here, we see the first hints of the separation between independent measurements and independent state preparations. This joint independence leads to the concept of statistical independence in classical and quantum mechanics \cite{Fewster:2016mzz,EmerGe_proj_2}.}. It is important to note that such a procedure is only possible by first determining the spacetime metric, which involves solving the Einstein equations with global initial conditions and integrating over the entire spacetime. Therefore, this local approximation is only feasible by considering the global structure of spacetime. This will be further discussed in Sec. \ref{sec:O-W_UQM}.

\subsection{Quantum Ontology}

The context of quantum mechanics makes everything much more interesting. In quantum mechanics, the primary elements of reality are not particles localized in spacetime but quantum states described by vectors in Hilbert space. These states can exist in superpositions and exhibit spatiotemporal non-local correlations, challenging the classical notion of independence. Adopting a quantum ontology thus requires rethinking the independence and interaction of parts. This perspective forces a reconsideration of how we define and measure the components of a system, as well as how we understand their interactions, topics we will explore in Sec. \ref{sec:why_care} and Sec. \ref{sec:O-W_UQM}. Quantum mechanics seems to suggest a more holistic view, where the properties of the whole system cannot always be decomposed into the properties of its parts. This reasoning also justifies approaches to quantum cosmology and views where reality is seen as an evolving vector in Hilbert space \cite{Carroll:2021aiq}.

\paragraph{Quantum Holistic Ontology} proposes that the fundamental nature of reality is best understood as an interconnected whole rather than as a collection of independent parts. This view is deeply rooted in the principles of quantum mechanics, where entanglement and superposition create correlations between systems that defy classical notions of separability. Unlike classical systems, quantum systems do not merely consist of discrete entities with localized interactions; instead, they are described by global quantum states in Hilbert space, which capture the complex, non-local relationships between parts of the system.

From this perspective, the main reason to consider this ontology is to attribute a sense of reality to entities that satisfy the Schrödinger equation, in contrast to quantum subsystems that seemingly demand some form of non-unitary evolution, apparent or not. Thus, the measurement problem will play a crucial role in how we approach this topic in Sec. \ref{sec:O-W_UQM}.

\paragraph{Atomistic$^{(K)}$ Quantum  Ontology} explores the possibility of maintaining a form of atomism within the framework of quantum mechanics, where the universe is composed of distinct quantum subsystems. Unlike classical atomism, which relies on spatial localization and independence, each quantum subsystem is associated either with factors in Hilbert space for finite-dimensional systems (Sec. \ref{sec:sapce_QM}) or by commuting local operator algebras in algebraic quantum field theory (Sec. \ref{sec:status_art}). However, as we will discuss in Sec. \ref{sec:GR_dres}, quantum gravity seems to undermine these decompositions, at least as long as they are understood as invariant under the unitary evolution of the global Hilbert space to which they belong. We will revisit this in Sec. \ref{sec:O-W_UQM}, where we will offer an alternative way to bridge holistic and atomistic quantum ontology, providing a new perspective on the measurement problem.

\subsection{How do we decide?} 

As mentioned briefly above, we seem to be trapped in spacetime with classical measurements. Thus, given our invocation of scientific realism, which employs rationales such as the no-miracles argument \cite{sep-scientific-realism} to attribute a sense of reality to theoretical terms in our theories, we soon are led into problems, such as our mysterious box in Sec. \ref{sec:mysterious}. There, we saw that under our limited measurements, only the energy displayed by the box, we could not infer the nature of the quantum system inside the box.

As discussed in Sec. \ref{sec:sapce_QM}, atomistic classical and quantum ontologies are intrinsically interconnected by the requirement that local measurements can be performed independently, establishing a direct connection between macro- and microcausality. Now, if there is ambiguity at the quantum level about how the Hilbert space is partioned and how local the quantum subsystems are interacting, we could imagine that further criteria should be imposed to select a preferred factorization of the Hilbert space, or assignment of local algebras. This project is pursued to some extent\footnote{Other ideas for selecting a preferred factorization of the Hilbert space have been recently pursued in \cite[e.g.]{Carroll:2020gme, Adil:2024wok}, with general discussions about decompositions of Hilbert space found in \cite{Niederklapfer}.} in \cite{Cotler:2017abq} where they argue that imposing a quantum mereology resulting in the most possible quantum-local Hamiltonian is (almost always) sufficient to select a particular TPS, which then leads to a specific classical limit. From the perspective of classical ontology, we can invert this picture and view their program (which could be thought of as establishing the grounds for discussing quantum mereology from a Hilbert space fundamentalist perspective) as recognizing that we indeed live in a specific classical limit and then finding the necessary criteria at the quantum level aligned with this classical limit. 

It is in this same vein that our approach to the measurement problem will be pursued in Sec. \ref{sec:O-W_UQM}. But first, we need to understand how the current state-of-the-art connects the quantum and classical realms, and then show how a quantum theory of gravity jeopardizes such a picture. 

%% file: status_of_art.tex
\section{Status-of-art, \emph{or crash course on algebraic quantum field theory}}\label{sec:status_art}

At the end of the day, we are seemingly trapped in spacetime. By this, we mean that the \emph{real patterns} \cite{Dennett1991-DENRP} surrounding us are best described by spacetime physics, which includes all the possible measurements we make \cite{Matsas:2007zz,Matsas:2023lmp}. In fact, despite their complexity, measurements still amount to simply counting—such as the number of clicks in a detector or spots on a screen—from which we can infer the state of our physical systems. 

Returning to the language introduced in Sec. \ref{sec:physical_theories}, the expectation values $\langle \Ocal \rangle$ of our measurements are a combination of a certain quantity $\Ocal$ given some state $s$ of the physical system:
\begin{equation}
    \langle \Ocal \rangle \equiv \langle s | \Ocal | s \rangle \,.
\end{equation}
You can think of each quantity as representing a real physical apparatus that is meant to make a specific measurement, with its display providing the range of possible values it can take once a system is measured.

The question is: how do we connect the fact that we are in spacetime, with our measurement devices and physical systems in our labs, to the more abstract formalism of quantum theory? In other words, how should we map the more fundamental quantum world into the realm of classical spacetime?

To understand this, we will qualitatively describe the state-of-the-art in our algebraic understanding of quantum theory. We will keep things as simple as possible to build the necessary intuition, allowing us to appreciate where things might go wrong as we approach the regime of quantum gravity. Additionally, we will rely on Sec. \ref{sec:sapce_QM} to discuss the interplay between the concepts of locality and physical systems across classical spacetime and quantum Hilbert space. By the end of our discussion, the mystery of the box from Sec. \ref{sec:mysterious} should become clearer, and we will have introduced the necessary tools to consider a new approach to the measurement problem in Sec. \ref{sec:O-W_UQM}.

\subsection{What could we have hoped for?}\label{sec:hope}

We operate under the belief that the world is fundamentally quantum-mechanical and that the entire spacetime is associated with an underlying global Hilbert space. Here we see the relevance and direct connection between the holistic ontologies discussed in Sec. \ref{sec:ontologies}. A crucial aspect of our activities in spacetime is the idea that it can be divided into independent parts. This concept underlies the separation between a system to be measured, our devices, and even ourselves as observers. The question then becomes what is needed to ensure that the whole can be broken into independent pieces, and to what extent this can be realized in Hilbert space. 

Unexpectedly, this is a deep philosophical and physical problem that also appears in many other disciplines. We do not intend to provide an exhaustive discussion of the matter, but we will highlight at least one criterion that we already met that is needed classically to define any reasonable notion of independence: \emph{causal independence}. If we know that two systems cannot exchange any information in a way that all possible observers would agree on, this is a strong criterion for considering them as independent entities. Special and general relativity provide the framework in which this can be formulated. We will refer to this causal independence as \emph{spatiotemporal locality}, as it establishes the condition under which we can securely say that something is independently located relative to something else (if they were not causally disconnected, there would be observers for whom the systems could be arbitrarily close to each other).

Let's be more precise. What we have in mind are two small regions of spacetime, $U$ and $V$, which are spacelike separated and define local Cauchy surfaces. This means that there is a spacelike surface inside each region that contains all the necessary information to describe anything inside those regions. 

Thus, what we could have hoped for is an analog of them in Hilbert space. That is, the entire Hilbert space can also be broken down into independent pieces, with each piece mapping directly into these regions in spacetime. This seems like a natural expectation: we have independent regions in spacetime fundamentally emerging from independent patches in Hilbert space, allowing us to precisely relate the prepositional verb \emph{to be in spacetime} to the verb \emph{to be in Hilbert space}. It turns out that this is not immediately possible, and the situation is aggravated when we attempt to include gravity in the picture. 

If we recall our initial mysterious box, perhaps this should not come as a surprise. Although quite often locality in Hilbert space, in the sense of local subsystems with commuting algebras, is a direct consequence of spatiotemporal locality (as discussed in Sec. \ref{sec:sapce_QM}), this is mostly our spacetime parochialism. As emphasized in recent works \cite[e.g.]{Maldacena:2013xja,Cotler:2017abq,Carroll:2020gme,Adil:2024wok}, the notion of spatiotemporal locality should be seen as emergent from the quantum formalism. So what can be done instead? 

\subsection{The Algebraic Approach}

There is, however, a direct mapping of spatiotemporal locality to another space: the \emph{space of observables}. Instead of trying to define smaller patches of Hilbert space associated with causally independent spacetime regions, we can associate a local algebra of observables with each region. These algebras commute with each other if the regions are causally independent,
\begin{equation}
    [ \Rcal(U), \Rcal(V)] =0\,,\label{eq:observables_QFT}
\end{equation}
where $U$ and $V$ are the two spacelike-separated regions. This is a defining property in algebraic quantum field theory (AQFT). AQFT starts by associating algebras of observables with regions of spacetime. For an open, finitely extended region \( U \) in Minkowski space, there is an associated algebra \( \mathcal{R}(U) \), where the self-adjoint elements of \( \mathcal{R}(U) \) are interpreted as observables measurable within \( U \). These algebras must satisfy several properties that capture the essence of locality and causality, among them \cite{Haag:1992hx}:

\begin{itemize}
    \item \textit{Isotony}: If \( U \subset V \), then \( \mathcal{R}(U) \subset \mathcal{R}(V) \);
    \item \textit{Microcausality}: If \( U \) and \( V \) are spacelike separated, then all observables in \( \mathcal{R}(U) \) commute with all observables in \( \mathcal{R}(V) \), i.e., \( [\mathcal{R}(U), \mathcal{R}(V)] = 0 \);
    \item \textit{Causality Structure}: Let \( \widehat{U} \) denote the causal completion of \( U \). The algebra associated with \( \widehat{U} \) is the same as the algebra associated with \( U \), i.e., \( \mathcal{R}(\widehat{U}) = \mathcal{R}(U) \).
\end{itemize}
These properties ensure that the theory respects the causal structure of spacetime, corresponding to the hyperbolic propagation character of fields. This is very similar to the algebraic structure present in finite-dimensional quantum systems reviewed in Sec. \ref{sec:sapce_QM}. To some extent, this is quite intuitive. Imagine you are inside one of these regions with your set of measurement devices. You know that they cannot influence the measurement devices of another causally disconnected region, as discussed earlier. Therefore, it seems reasonable to imagine that the operators associated with your measurement devices commute with those from other causally disconnected spacetime regions. 

However, note that the operators in the equation above are not defined over smaller patches of the Hilbert space; instead, they are global operators. So, flipping things around, imagine the entire Hilbert space associated with the entire spacetime\footnote{\label{ft:states}This is a possibly contentious statement. In AQFT, different vacua lead to inequivalent representations of the algebra of observables, resulting in different Hilbert spaces \cite{fewster2019algebraic}. In fact, these inequivalent representations arise from various superselection sectors and symmetry-breaking patterns, fundamental for understanding phenomena such as spontaneous symmetry breaking in physics. An example is the different vacuum states in gauge theories, leading to distinct physical sectors that cannot be coherently superposed. However, we expect that this is not a problem from the point of view of a final, unified theory of nature.}. Then, consider all the possible observables in it and start separating them into commuting subalgebras. So far, this has nothing to do with spatiotemporal locality. However, once you start labeling these subalgebras with \( U \), \( V \), etc., you can use that as a condition to establish spatiotemporal locality as described before. Thus, spatiotemporal locality might emerge from commuting subsets of observables defined across the entire Hilbert space. 

But with what states should these observables be paired? As we discussed in Sec. \ref{sec:physical_theories}, relevant quantities yield values that can be related to measurements only when combined with states. Similarly, we can define a state space over the whole algebra, thus they are global states, but that can have their domain restricted to the local subalgebras. This means that when combined with the local algebra of observables \( \mathcal{R}(U) \), they generate values corresponding to local measurements. It is important to distinguish this from the concept of a smaller Hilbert space, and its respective local states, given the obstruction discussed in Sec. \ref{sec:hope}.

Here we arrive at a crucial question: how do we know that the values generated by combining these global operators and states have anything to do with the values we actually measure in labs in spacetime? Well, that is exactly what modeling entails. We theorize quantum systems to reproduce what we observe in the laboratory. However, we remain trapped at the upper level in spacetime, and it is worth asking whether there could be other quantum models that would lead to the same measured values in spacetime, much like our mysterious box.

By analogy with coarse-grained emergent descriptions discussed in Sec. \ref{sec:emergence_lvl_0} and \ref{sec:emergence_lvl_1}, in the space of quantum models aimed at reproducing classical spacetime measurements, is there a notion of an `ensemble' of models that all converge to the same set of numbers one obtains by taking measurements in the lab? Is there a notion of model coarse-graining that needs to be considered because, similar to our accessible macrostates in thermodynamics, we only have access to classical measurements in the end? We will return to these questions in Sec. \ref{sec:O-W_UQM}, but we have already seen that in the case of dualities, this is exactly the situation: different sets of commuting subalgebras can be used to arrive at the same classical measurements, even though they give rise to different spatiotemporal local structures, as observed in Sec. \ref{sec:sapce_QM}. 

This highlights the subtleties involved in defining quantum subsystems and the emergent nature of spatiotemporal locality from a more fundamental quantum framework. And now, at least we finally understand why the maps between the different sets of spin sites were non-locally related, since the choice of commuting subalgebras is one of the criteria that leads to the notion of independent subsystems. 

\subsection{Trapped in Spacetime}\label{sec:trapped_ST}

In the previous section, we discussed how the algebraic approach in AQFT allows us to map spatiotemporal locality to the space of observables. This highlights how our classical understanding of locality and independence can emerge from a more fundamental quantum framework. However, this reliance on classical concepts is not merely a convenient tool; it is deeply ingrained in the very formulation of quantum theory.

It has long been realized by some of the founders of quantum mechanics that the theory is rather particular, especially in the context of ``quantization," where we impose our classical preconceptions, such as spatiotemporal locality, onto the quantum world. In fact, Bohr viewed this as an unavoidable necessity:
\begin{quote}
    ‘It is decisive to recognize that, however far the phenomena transcend the scope of classical physical explanation, the account of all evidence must be expressed in classical terms. The argument is simply that by the word experiment we refer to a situation where we can tell others what we have done and what we have learned and that, therefore, the account of the experimental arrangements and of the results of the observations must be expressed in unambiguous language with suitable application of the terminology of classical physics.’ (Bohr, 1949, p. 209 \cite{Bohr1949-BOHDWE}) 
\end{quote}
Similarly, Heisenberg says: 
\begin{quote}
    ‘The Copenhagen interpretation of quantum theory starts from a paradox. Any experiment in physics, whether it refers to the phenomena of daily life or to atomic events, is to be described in the terms of classical physics. The concepts of classical physics form the language by which we describe the arrangement of our experiments and state the results. We cannot and should not replace these concepts by any others.’ (Heisenberg 1958, p. 44 \cite{heisenberg2007physics}) 
\end{quote}
Both argue prescriptively that classical physics is not going anywhere, but rather is fundamentally necessary in the context of measurements in quantum mechanics. A few decades later, Landau and Lifshitz summarize the situation as: 
\begin{quote}
    ``Thus quantum mechanics occupies a very unusual place among physical theories: it contains classical mechanics as a limiting case, yet at the same time it requires this limiting case for its own formulation.'' (Landau \& Lifshitz, 1977, p.3 \cite{landau1977quantum})
\end{quote}

These quotes have been beautifully compiled in the introduction chapter of \cite{landsman2017foundations}, where the author himself provides a more rigorous formulation of such a prescription in terms of our modern understanding of the algebraic approach to quantum mechanics:
\begin{quote}
    The physically relevant aspects of the non-commutative operator algebras of quantum-mechanical observables are only accessible through commutative algebras. \\(Landsman, 2017, p. 10)
\end{quote}

Now that we understand how classical notions of spacetime are intrinsically embedded in quantum theory, it is time to extend our discussion to a dynamical spacetime and its interaction with the quantum framework. As general relativity can be seen as a field theory with long-range interactions,  we will first review how gauge theories are treated in the context we have been discussing to ease our way into this complex discussion.

\subsection{More ambiguities!}\label{sec:QED_scalar}

We have been discussing the quantum-mechanical observables associated with microcausality since the beginning of this paper, but we have not yet addressed what they are in the context of field theory. Unlike in finite-dimensional quantum mechanics, where the simple assumption that two quantum subsystems are spacelike separated is enough to represent their set of possible observables as two commuting sets of local algebras, and where their interacting Hamiltonian can be thought of as something we switch on and off in the lab, in quantum field theory, we do not have such luxury. Instead, we have long-range interactions that are ever-present and are modeled by gauge theories, where the observables associated with the free theory typically transform under gauge transformations. Thus, even if the free theory would allow us to define commuting sets of subalgebras associated with spacelike-separated regions, thereby satisfying \eqref{eq:observables_QFT}, once the gauge interaction is taken into account, that microcausal relationship is no longer gauge invariant.

Since we do not observe violations of macrocausality in the lab, and macrocausality is intrinsically connected to microcausality, the question becomes whether our quantum field theories allow the construction of both gauge-invariant and microlocal observables. The expectation is that those are the ones actually measured in the lab. Let’s explore how this works in quantum electrodynamics (QED) \cite{Donnelly:2015hta} as an example. 

\subsubsection*{QED couple with scalar field}

Let's consider QED coupled with a scalar field $\phi$ with charge $q$,
\begin{equation}
    \mathcal{L} = -\frac{1}{4}F_{\mu\nu}F^{\mu\nu}-\frac{1}{2\alpha}(\partial_\mu A^\mu)^2 -|(\partial_\mu -iqA_\mu)\phi | ^2 - \frac{1}{2}m^2 |\phi|^2, 
\end{equation}
where different choices of $\alpha$ fix the gauge: for instance, $\alpha\rightarrow0$ corresponds to Lorenz gauge. Being a gauge interaction, the gauge field transforms, as well as any other charged field under it,
\begin{align}
 A_\mu \to A_\mu - \partial_\mu \Lambda\quad \text{and} \quad \phi(x) \to e^{-\mathrm{i} q \Lambda(x)}\phi(x)\,,
\end{align}
such that the Lagrangian is invariant, as gauge transformations are ambiguities in the description of the gauge interaction, and physics should not depend on them. 

If we could turn off the gauge interaction, or effectively deactivate the scalar field charge, we could consider the complex scalar field components as a local observables (assuming some smearing) associated with different spacetime regions, such that for any two spacelike-separated regions \(x \in U\) and \(y \in V\), microcausality holds:
\begin{equation}
    [\phi(x), \phi(y)] = 0\,.
\end{equation}
This makes sense since the field itself, acting on the vacuum, can create local excitations associated with particles, which certainly seem to be observed in the lab. However, since we cannot simply turn off the interaction, this field operator is not gauge-invariant, raising the question of whether the particle content observed in the lab is well-posed by the theory.

Fortunately, we can \emph{dress} the field observable in a way that makes it gauge-invariant by defining a different observable:
\begin{equation}
    \Phi(x) = V(x) \phi(x)\,,
\end{equation}
where the dressing \(V(x)\) is defined such that it takes the inverse transformation of the field, \(V(x) \to e^{iq\Lambda(x)}V(x)\), making \(\Phi(x)\) gauge-invariant. Such a dressing is found to be:
\begin{equation}
    V(x) = \exp \left( iq \int d^4x' f^\mu(x,x')A_\mu(x') \right)\,, \quad \text{where} \quad \partial_\mu' f^\mu (x,x') = \delta^4(x-x')\,, \label{eq:QED_dressing}
\end{equation}
assuming the gauge transformations vanish at infinity. There is considerable freedom in choosing the function \(f^\mu\), and different choices lead to different dressings. The physical intuition behind this dressed observable is that it creates particles of the scalar field along with a \emph{cloud} of photons surrounding each particle. Different dressings correspond to different configurations of these clouds.

Having found gauge-invariant observables, the question turns to whether they are microcausal. Upon quantizing this theory, one imposes canonical commutation relationships for both fields. For spacelike-separated regions, we have:
\begin{equation}
    [\Phi(x), \Phi(y)] = \phi(x) \phi(y) [V(x), V(y)] + V(y) V(x) [\phi(x), \phi(y)]\,, \label{eq:QED_dres_micro}
\end{equation}
which vanishes despite the non-local nature of \(V(x)\). This can be most easily seen when one picks a frame where it becomes an equal-time commutator, and then any pair of commutators among the fields vanishes\footnote{We refer the reader to \cite{Donnelly:2015hta} for more details.}. Despite \(\Phi(x)\) being gauge-invariant and defining local algebras that satisfy microcausality, it is intrinsically a non-local operator as the integration extends to infinity. However, another algebra of operators can be constructed in which we have local operators instead, where the electromagnetic field is screened by introducing a second operator creating an oppositely charged particle \cite{Donnelly:2016rvo}:
\begin{equation}
    D(x,x') = \phi(x) e^{iq\int_\Gamma dx^\mu A_\mu} \phi^*(x')\,,\label{eq:QED_local_dress}
\end{equation}
where $\Gamma$ is a curve connecting $x$ and $x'$ and it represents a Wilson line connecting two field excitations of opposite charges. Then, as long as $D(x,x')$ can be localized in $U$, it will commute with analogous operators defined in other spacelike-separated regions, and once again we find a gauge-invariant algebra of observables that is microcausal.

%% file: gravity.tex
\section{What about gravity?}\label{sec:gravity}

Any good story starts with a mystery and, at some point, introduces a villain. Our understanding of gravity has dramatically changed since Newton's theory. Initially, space and time were deemed absolute, and gravity was yet another force acting on this absolute stage. However, we have since learned with Einstein that the stage itself is dynamical, changing depending on the matter-energy act at play, and this very dynamics is responsible for gravitational phenomena.

One of the key principles guiding Einstein's development of his theory of gravity was \emph{relationalism}, which can be understood as \cite{Wuthrich2024}:
\begin{quote}
    Space and time do not exist as independent substances, there is only the material content of the universe. Space and time are merely defined through spatiotemporal relations among the material objects in the universe (and their parts). Spacetime is thus a (constructed) relational complex of relative positions of material objects. (Wüthrich, 2024)
\end{quote}
We can directly appreciate his take on it when reflecting on the development of relativity \cite{lehmkuhl}:
\begin{quote}
    The systematic decoupling of basic geometrical concepts (straight
    line, distance, etc.) from the bodies of experience, of which they
    are abstract representations, must not let us forget that in the
    end geometry is supposed to tell us about the behavior of the
    bodies of experience. If there were no practically rigid bodies that can be brought into congruence with one another, we
    would not speak of the congruence of distances, triangles, etc. It
    is clear that for the physicist geometry becomes meaningful only
    as he associates bodies of experience with those basic concepts,
    for example by associating the concept of distance with a practically rigid body with two markings. Vice versa, this association
    makes Euclidean geometry a science of experience in the truest
    sense, just like mechanics. The sentences of geometry can then
    be confirmed or falsified, just like the sentences of mechanics. (Einstein, 1919-20)
\end{quote}
Thus, it is quite clear that Einstein's perspective does not take space, and perhaps time, as independent substances. However, Einstein's general relativity did not quite perfectly achieve this strict version of relationalism, as it does allow vacuum solutions, but it was a significant step in that direction. Regardless, general relativity is extremely successful and underlies both astrophysics and modern cosmology, each presenting a large body of evidence towards its validity across different physical regimes.

However, it is not clear whether general relativity applies to all regimes. In fact, most theoretical evidence points to its breakdown, including the theory's own solutions in the context of spacetime singularities, such as black holes or anything resembling the beginning of the Universe. On one hand, physicists are not overly concerned, as it is believed that general relativity is simply a limit of an underlying quantum theory of gravity where such issues would not be present; on the other hand, we are approaching the turn of the century since the first attempts to quantize gravity  \cite{rovelli2001notes}, yet there is no clear consensus on the correct path to take. One reason for this lies in the very different nature of quantum mechanics and general relativity, as highlighted here \cite{Hohn:2017cpr}:

\begin{quote}
   (Quantum) physics lies in the relation between the observer and observed systems; one can
regard quantum theory as a law book which governs the observer’s acquisition of information
from physical systems in terms of \textit{how} and \textit{how much} information is accessible. In quantum
theory the information flow is physical, encoded in physical systems that can undergo their own
dynamics, such as electrons. This physical information flow is under the focus of the theory but
the spatiotemporal stage on which such information flows is externally given.

By contrast, (...) much of the physics of general relativity lies in the
communication relations among observers (or more generally systems); general relativity can
be understood as a law book governing \textit{where} and \textit{when} information can be communicated. In
general relativity the information flow among idealised observers, at least in terms of light signals,
is, strictly speaking, external, while the stage on which this information flows, i.e. spacetime, is
dynamical and the focus of the theory. Alice can send a light signal to Bob, depending on their
causal relations, but the influence of this light signal on the dynamical spacetime they live in is
ignored, no matter how energetic it is. (Höhn, 2017)
\end{quote}

Nonetheless, many lessons have been learned so far. But we will focus on one lesson that will lead us to reconsider several approaches to important problems in different areas of physics in Sec. \ref{sec:why_care}. This is the problem of defining local gauge-invariant observables in linearized quantum gravity. 

\subsection{Operational Measurement Limitation}

Let's start intuitively \cite[see interview with Nima Arkani-Hamed]{armas2021conversations}. Imagine you are back in your lab, and you want to make a perfectly sharp measurement, or at least as sharp as possible. The problem is how to store that information in your apparatus. Quantum mechanically, you could imagine having an infinite chain of qubits to record any measurement using binary digits. This would take a lot of space, so perhaps you have access to a quantum field that has an infinite-dimensional Hilbert space at each point in space. Either way, you need an infinite-dimensional Hilbert space for a perfectly sharp measurement. While this may not be feasible in practice, there is nothing in quantum mechanics that prevents you from imagining it in principle.

However, once gravity is considered, all this stored information will eventually start to backreact on the geometry of the lab, until it reaches a threshold and transforms it into a black hole. This is not surprising, as black holes are the most information-dense objects according to the Bekenstein bound \cite{Bekenstein:1980jp,Bousso:1999xy,Bousso:2002ju}. Since the entropy of a black hole scales as \( S = \frac{A}{G} \), where \( A \) is its area, and it is a maximally entropic state, we can infer that the Hilbert space associated with it has a dimension of approximately \( \mathcal{H} \sim e^S \) \cite{Bao:2017rnv,Cao:2016mst}. Thus, the error in the experiment will scale as \( e^{-A/G} \). Therefore, there is no precise observable one can associate with any measurement performed in a finite-sized room.

Is there a way out of this? One approach to bypass these constraints is to imagine a lab that can extend indefinitely instead of having a fixed size. As you record more digits from your measurements, you avoid saturating the local bound on information density and prevent creating a black hole. However, to keep sharpening the measurement, you will eventually need to push your apparatus all the way to the boundary of spacetime. This is why the only sharp observables in gravity are those associated with the boundary. In asymptotically flat spacetime, it is the S-matrix that defines such observables, with in and out states defined at infinity. Similarly, in black hole physics, significant quantities are often defined globally.

Now, let's explore more concretely what happens when we try to define local gauge-invariant observables in gravity.

\subsection{\emph{You can’t always get what you want}}\label{sec:GR_dres}

Let's now consider general relativity coupled with a scalar field and pursue a similar strategy as we did in Sec. \ref{sec:QED_scalar} for QED coupled with a scalar field \cite{Donnelly:2015hta}. Thus, our starting point is
\begin{align}
\mathcal{L} = \frac{2}{\kappa^2} R - \frac{1}{2} \left( g^{\mu\nu}\nabla_\mu \phi \nabla_\nu \phi + m^2 \phi^2\right)
\end{align}
where $\kappa = \sqrt{32\pi G}$. This theory is diffeomorphic invariant. We will consider a fixed flat background with perturbations on top, so we expand $g_{\mu\nu} = \eta_{\mu\nu} + \kappa h_{\mu\nu}$, where  $h_{\mu\nu}$ is a canonically normalized field. Under infinitesimal linearized diffeomorphisms, the fields transform as 
\begin{align}\label{eq:gauge_trafo}
\delta \phi = - \kappa \xi^\mu \partial_\mu \phi + \mathcal{O}(\kappa^2) \quad \text{and} \quad \delta h_{\mu \nu} =- 2 \partial_{(\mu} \xi_{\nu)} + \mathcal{O}(\kappa)
\end{align}
for an arbitrary vector field $\xi^\mu$. Once again, the scalar field alone is not invariant and cannot be used to define local-diff invariant observables. But looking at its transformation, in a fixed background, the linear diff transformation of the scalar field corresponds to a local infinitesimal translation. As we know how to represent translations of any observable by a unitary transformation, the operator that we seek takes the form 
\begin{align}
\Phi(x) = e^{\mathrm{i} V^\mu P_\mu} \phi(x) e^{-\mathrm{i} V^\mu P_\mu} = \phi(x^\mu + V^\mu)
\end{align}
where $P^\mu=-i\partial_\mu$ is the translation operator and $V^\mu$ is a functional of the metric perturbation, which at linear order we require to transform as 
\begin{align}\label{eq:delta_V}
\delta V^\mu = \kappa \xi^\mu \,. 
\end{align}
Using \eqref{eq:gauge_trafo} and \eqref{eq:delta_V}, it is straightforward to show the invariance of $\Phi$ at leading order in $\kappa$. As we will discuss more in Sec. (\ref{sec:why_care}), we expect that any final theory of quantum gravity will have an effective perturbative regime that falls into the theory being studied here, and we argue that much of the current physical phenomenology is modeled, tacitly or explicitly, under this linear perturbative regime.

The real question is what $V^\mu$ is and how it depends on the metric. We could imagine that we should seek its form similar to what we did in \eqref{eq:QED_dressing}, 
\begin{equation}
    V^\mu(x) = \kappa \int d^4x'f^{\mu\nu\lambda} (x,x') h_{\nu\lambda} (x')\, \quad \text{where} \quad 2\partial_\nu'f^{\mu\nu\lambda}(x,x') = \delta^4(x-x')\eta^{\mu\lambda}\,,
\end{equation}
as it does guarantee that $V^\mu(x)$ transforms as \eqref{eq:delta_V}.  However, it turns out that $f^{\mu\nu\lambda}$ depends non-locally on the metric. And that should have been expected, because in \eqref{eq:QED_dressing}, it was the gauge field that was used for the dressing, and the gauge field in GR is not the metric but the connection. To see how this impacts the dressing, let's take a look at a single example of $V^\mu$ associated with the gravitational Wilson line \cite{Donnelly:2015hta,Giddings:2019hjc,Giddings:2018cjc}:
\begin{align}
V_\mu = \frac{\kappa}{2} \int_x^\infty d x'^\nu \left[h_{\mu\nu}(x') + 2 \int_{x'}^\infty d x''^\lambda  \partial_{[\mu} h_{\nu] \lambda}(x'')\right]\,,
\end{align}
where the integration is along an arbitrary curve from $x$ to infinity.
In physical terms, the dressing corresponds to a gravitational string extending to infinity. The crucial difference between the gravitational dressing and the one in QED, and more generally Yang-Mills theories, is the presence of the derivatives of the metric.

Now that we have obtained dressed gravitational operators that are invariant under linear diffs, do they satisfy microcausality? Unfortunately, not. As we had before, the commutator of $\Phi$ between two spacelike regions depends on $[V(x),V(y)]$ as in \eqref{eq:QED_dres_micro}, but inside of $V^\mu$ there is the metric and its derivatives, including its time-derivative. What was crucial for the QED example was that the dressing was made with the gauge field, which is the same object that was quantized, but that is not the case for gravity. Typically, in the covariant quantization of gravity, both non-perturbative and linearized gravity, we quantize the metric instead, which its necessity can be traced back to the introduction of the Ricci scalar (both the Riemann tensor and Ricci tensor can be fully written only in terms of the connection)\footnote{Here, we see a very clear motivation to consider Loop Quantum Gravity where the connection, expressed in terms of Ashtekar variables, is quantized instead of the metric. For a recent discussion, \cite[and references therein]{Rovelli:2022xsj}.}. You might wonder about the other type of dressed observable \eqref{eq:QED_local_dress}, but those are not available due to the absence of negatively charged particles under gravity. 

The evidence against the existence of gauge-invariant local observables in quantum gravity is compelling. Already three decades ago, \cite{Torre:1993fq} demonstrated that in closed universes, invariant observables cannot be constructed as local functions of the canonical variables, linking the absence of local observables to topological symmetries. More recently, \cite{Marolf:2015jha} argues that while classical gravitational observables can be microcausal, this property is limited at quantum scales, with significant implications for black hole information. Additionally, \cite{Donnelly:2016rvo} shows that any operator with nonzero Poincaré charges must be gravitationally dressed, implying nonlocality and the failure of defining local subsystems in gravity and \cite{Giddings:2018cjc} highlights the conflict of defining mutually independent subsystems via standard tensor products or operator algebras in gravitational contexts. For more references, see \cite[section \textit{Observables} and references therein]{deBoer:2022zka} and \cite{EmerGe_proj_1}.

%% file: why_care.tex
\section{Why should we care?}\label{sec:why_care}

Despite all the candidates, we still have not reached a consensus regarding the theory of quantum gravity. However, it is widely accepted that any theory of quantum gravity must have a low-energy quantum gravity (LEQG) limit, which at energies well below the Planck scale, and recovers the effective field theory derived from general relativity by quantizing metric perturbations on a classical background. This argument is extensively developed in \cite{Wallace:2021qyh}, where it is also shown that the body of evidence from astrophysics and cosmology, which requires a simultaneous treatment of quantum-mechanical and gravitational effects, necessitates such a theory.

More recently, the prospect of probing the quantum nature of gravity in the laboratory has been proposed \cite{Marletto:2017kzi, Bose:2017nin}. It is claimed that detecting entanglement between two mass probes, mediated by gravitational interaction, would demonstrate the quantum nature of gravity \cite{Chitambar:2014svb}. The modeling of such experimental proposals makes direct use of LEQG, to the extent that some argue that the aspects of quantization these experiments can probe are the same ones already probed in cosmological observations \cite{Fragkos:2022tbm}.

However, in light of our discussion so far, we should not expect to be able to measure gauge-invariant observables in LEQG that also respect microcausality. We have seen that microcausality underpins our operational notion of macrocausality. Thus, in regimes where LEQG is expected to play a major role, both microcausality and macrocausality violations might occur. For instance, macrocausality in the early universe is a key assumption to introduce the horizon problem that led to the inflationary paradigm. Meanwhile, in Sec. \ref{sec:sapce_QM}, we established that microcausality, and more generally having commuting sets of observables for independent quantum systems, is a necessary condition to have a well-defined tensor product structure. This structure underlies the notion of quantum-mechanical systems and is tacitly assumed in the formulation of the measurement problem in quantum mechanics, and indeed in any measurement, including those trying to probe the quantum nature of gravity in the lab.

In this section, we briefly review physical regimes where LEQG is significant, highlight assumptions of microcausality and macrocausality, and discuss potential future directions.

\subsection{Early Universe Cosmology}

We do not have direct access to the very early universe, but among all the different cosmological observations that offer us a window into its early stages, the Cosmic Microwave Background (CMB) radiation takes the main role. Spread out throughout the whole universe, it provides an aged snapshot of the Universe approximately 380,000 years after the Big Bang. It is a snapshot in the sense that this background radiation was produced during a short phase where the light in the early Universe went from propagating in an opaque medium, where the photons were constantly interacting with electrons and baryons, to a transparent one where they could freely propagate until reaching our current detectors. It is aged because the Universe has evolved since this transition, and the detected radiation distribution today carries memories of the Universe's history when compared to its expected profile back when it was emitted. For example, one of the memories impinges upon the whole radiation spectrum, which has been redshifting since its emission, and today is detected at a temperature of $2.7$ K as opposed to its initial temperature of around $3000$ K. 

As we discussed in Sec \ref{sec:emergence_lvl_1}, if we are talking about temperature, it must mean that we are close to thermal equilibrium. That is the case for the CMB, which presents an overall spectrum very close to that of a black body, with small temperature fluctuations of order $10^{-5}$ K. These fluctuations are also part of the spectrum’s memory, as they reflect small density variations in the early Universe. One of the initial surprising features of the CMB was exactly this proximity to thermal equilibrium. The medium transition from opaque to transparent happens as the local temperature decreases enough such that, on average, photons cannot excite electrons out of the proton’s electric potential anymore, allowing them to form stable hydrogen. This local temperature depends on the amount of local energy density. Observing thermal equilibrium in the CMB indicates that the energy density was very evenly distributed both homogeneously and isotropically; otherwise, different parts of the Universe would reach this transition at different times, leading to larger variations of the temperature of the detected photons depending on which part of the Universe we are probing (or, literally, looking at). This even distribution of energy density is particularly puzzling because, under the standard Big Bang evolution, the CMB corresponds to causally disconnected patches of the universe. Therefore, we need to explain how the Universe’s energy distribution was so even at the time the CMB was emitted\footnote{For more quantitative details, see \cite{mukhanov2005physical,baumann2022cosmology}.}.

Moreover, even if we could argue that the initial state of the Universe corresponded to perfect homogeneity due to unknown properties of quantum gravity, we would still need to explain why the fluctuations in the CMB are not random, but instead have correlations over very large distances that were not causally connected, so the light did not have enough time to travel that much since the beginning of the Big Bang. In fact, without assuming any new physics between the Big Bang and the emission of the CMB, we can show that points separated by more than two degrees on the sky do not seem to have been in causal contact. 

Therefore, to explain this amount of uniformity in the CMB many different ideas have been introduced, and today the early universe's paradigm is based on cosmic inflation: an early-on accelerated phase of expansion of the Universe that effectively blows up the size of the past causal horizon so that there would be enough time for light to propagate over large distances and create the sort of correlations observed in the CMB. The success of inflation comes from providing a mechanism for the generation of primordial density perturbations: they are quantum fluctuations that during inflation are stretched out to macroscopic scales, and they match the statistical properties of the temperature fluctuations in the CMB.

Thus, currently, the early universe paradigm is based on apparent violations of macrocausality due to a high degree of homogeneity of the radiation in the CMB together with seemingly super-causal correlations. Meanwhile, the current paradigm offers a solution based on inflation, which is typically modeled by a scalar field sourcing the fast-expanding cosmological background, while the combined quantization of the field and spacetime fluctuations is ultimately responsible for the primordial density fluctuations. Hence, the paradigm falls into LEQG.

This can potentially introduce some internal inconsistency. The reason is that the very reason to introduce new physics in the early universe is based on assumptions (macrocausality) that are not a given within the background theoretical framework (LEQG), as we discussed in Sec. \ref{sec:GR_dres}. 

Based on this argument, a future direction is to study the early universe using non-local gauge-invariant observables dressed by gravitational gauge fields. It is known that dressing a field theory creates a new Fock space from which scattering amplitudes can be analyzed; in particular, the states are defined over a new vacuum.  Thus, basically we have either a set of observables with the undressed vacuum or a dressed set of observables with the dressed vacuum. This 
is reminiscent of the setup of the Unruh effect, or Hawking radiation, where different observers, one inertial and one accelerating, have access to different sets of observables and vacua, and the vacuum state of one is seen as a thermal bath to the other. Could our observations of the CMB seen from the local perspective\footnote{Since gravity is not modelled in the detection, it is as if we are framing the observations from the point of view of the undressed theory.} look thermal exactly because of the mismatch between these vacua?\footnote{A similar argument can be made in the context of the cosmological constant problem, as the calculations making the problem explicit typically rely on the Fock space defined by the free quantum field theory \cite{Martin:2012bt}. It would be interesting to check whether the zero-point energies of the dressed modes contribute differently than the ones defined by the free theory.} 

\subsection{Gravitationally Induced Entanglement (GIE) Experiments}

The quantum nature of gravity remains a topic of debate \cite[and references therein]{Oppenheim:2018igd, oppenheim2023time}. Thus, it seems more than justified the hype around the possibility of probing whether gravity is non-classical in the lab \cite{Marletto:2017kzi,Bose:2017nin}. Originally, the experimental setup involved two masses, each in a spatial superposition, interacting only through gravity. Due to their spatial superposition, each pair of trajectories induces a different unitary phase due to the interaction Hamiltonian, which depends on the relative distance of the trajectories, different for all the pairs. Thus, if the initial state was a product state between the superimposed spatial states of each mass, after their interaction, each term will have a different phase, and the overall state can no longer be decomposed as a product state. Following \cite{Bose:2017nin}, if the initial state is considered
\begin{equation}
    |\Psi (t=0)\rangle = \frac{1}{2}(| L \rangle_1 + | R \rangle_1) (| L \rangle_2 + | R \rangle_2) \,,
\end{equation}
where $L$ and $R$ stand for left and right trajectories, then the final state is 
\begin{equation}
     |\Psi (t=t_f)\rangle = \frac{1}{2}\left( e^{i \phi_{LL}} |L \rangle_1 |L \rangle_2 + e^{i \phi_{LR}} |L \rangle_1 |R \rangle_2 + e^{i \phi_{RL}} |R \rangle_1 |L \rangle_2 + e^{i \phi_{RR}} |R \rangle_1 |R \rangle_2\right)\,.
\end{equation}
For example, $\phi_{LL} = Gm_1m_2t_f/\hbar (x_{L_1} - x_{L_2})$ represents the phase difference for the left-left trajectory. Therefore, the final state is entangled. This is a prediction of standard quantum mechanics. 

The experiment described above is known as the bipartite model, which is typically considered a Newtonian approximation of the tripartite model. For this model, gravity appears as a mediator subsystem, such that the masses interact only directly with it. Thus, for instance, instead of $|L \rangle_1|L \rangle_2$, we have $|L \rangle_1 |\gamma_{LL}\rangle |L \rangle_2$ which evolves to $e^{i \phi_{LL}} |L \rangle_1 |\gamma_{LL} \rangle |L \rangle_2$. As each of the $|\gamma\rangle$ states represents a different classical gravitational configuration, there is a superposition of classical gravitational states. What is crucial to the tripartite model assuming interactions between the masses only mediated by the subsystem $|\gamma\rangle$ is that, if the $|\gamma\rangle$ is a classical subsystem (the set of its observables define a commuting subalgebra), then one can show that no entanglement can be produced \cite[see for an overview]{Huggett:2022uui}\footnote{Previous arguments relied on LOCC-style reasoning \cite{Chitambar:2014svb}, typically assuming finite-dimensional systems. Only recently has this gap been closed by showing that any classical mediator, modeled as a commutative unital $C^*$-algebra, cannot generate entanglement, even for infinite-dimensional quantum systems and general $C^*$-tensor products \cite{Ludescher:2025kko}.}. Hence, were the experiment performed, witnessing the entanglement between the two masses would be strong evidence of the quantum nature of gravity. 

At least that is the common lore. Critical analyses of the modeling and implications of a positive detection have already been discussed \cite{Anastopoulos_2022}. In \cite{Fragkos:2022tbm}, it has been argued that such experiments are probing the quantum nature of gravity similarly to the quantized cosmological perturbations being probed by the CMB observations under the inflationary paradigm discussed above. They highlight two key differences: i) cosmological perturbations focus on field fluctuations, whereas GIE experiments involve cat states, which are significantly more quantum (measured by non-classicality such as Wigner negativity), and ii) GIE experiments explore a new, untested regime of physics, namely, the gravitation of delocalized masses.

Here, we want to call attention to yet two other aspects underlying the modeling of such experiments that have not been pointed out so far. Due to the low-energy regime and reliance on perturbative quantum gravity, these experiments fall within the LEQG framework. However, as discussed in Sec \ref{sec:sapce_QM}, a well-defined tensor product structure, necessary for composing different systems, is only ensured in the presence of microcausality. Moreover, the mass states above are also defined as excitations of some free matter field \cite[see supplementary material]{Bose:2017nin}, which, given our discussion in Sec. \ref{sec:GR_dres}, are not gauge invariant under linear diffeormorphisms.  Thus, the consistent use of LEQG to describe such experiments, believed to be fundamental to any possible conclusion to be derived from either positive or negative entanglement detections, seems to require an improvement of their description both in terms of the factorization understanding of the Hilbert space as well as the type of states that are considered\footnote{There is yet another issue that has to do with state preparation. In the context of field theories, the idea behind a product state is generalized to what is called statistical independence \cite{Fewster:2016mzz,EmerGe_proj_2}, which guarantees that any pair of state preparations made by two different experimenters in two spacelike regions can be seen as a preparation of the system as a whole. Achieving statistical independence requires both microcausality and the split property, which together ensure a well-defined tensor product structure and thus product states. Once again, LEQG might seem to violate microcausality, and it remains unclear whether quantum gravity would satisfy anything like the split property \cite{Donnelly:2017jcd,Donnelly:2018nbv,Raju:2021lwh}.}.

\section{Single-world unitary quantum mechanics}\label{sec:O-W_UQM}

Our discussion about the GIE experiments invites us towards a much broader shift in our conceptual perspective. The breakdown of microcausality in LEQG, and the subsequent loss of a well-defined tensor product structure, undermines the concept of quantum-gravitational subsystems, which, due to the universal nature of gravity, applies to everything. This shift may offer a new perspective on the measurement problem, suggesting that quantum subsystems are not invariant under measurements or generic Hamiltonian evolution\footnote{This idea was first explored in \cite{Boulle2023} by modeling an apparatus and a quantum system, both as qubits. Although the mathematical implementation was limited, some of the conceptual framework was developed there, though not in the context of LEQG and microcausality violations. More recently, thanks to Jan Glowacki and Tein van der Lugt, I became aware of \cite{ormrod2024quantum}, where similar ideas are pursued over solid mathematical foundations. Another paper that explores similar ideas can be found in \cite{Stoica:2024nff}.}. Just as the relativization of time and space was crucial for the development of special relativity, the relativization of quantum subsystems might be essential for resolving the measurement problem and advancing towards a final theory of quantum gravity. 

To some extent, this shift should not come as a surprise. A central lesson of general relativity is background independence: the idea that no structure should be fixed and exempt from dynamics. But a fixed tensor product structure in Hilbert space\footnote{Admittedly, our discussion so far has overlooked a crucial subtlety, particularly evident in the Heisenberg picture, where time evolution under the Hamiltonian naturally generates a continuous family of inequivalent TPSs \citep{Stoica:2021rqi,DiBiagioFranzmannSoulas}. In what follows, we will work with an initially fixed, kinematically selected TPS defined at an initial reference time \citep{Zanardi:2004zz}. Still, a more robust and general approach treats subsystems not in terms of Hilbert space factorization, but rather through the algebraic structure of operator subalgebras and their interrelations, an idea that finds its most precise expression in the formalism of descriptors \citep{Deutsch:1999jb,B_dard_2021}.}, or a fixed assignment of algebras to spacetime regions, violates this very principle. Once gravity is subject to quantum principles, it seems natural that such structures must also become dynamical. In fact, just as different spacetime foliations in special relativity are induced kinematically by different inertial observers, the kinematical relativization of subsystems in quantum theory was already emphasized in \cite{AliAhmad:2021adn}. What remains open is how to promote this relativization to a dynamical feature of the theory. Here, we aim to outline what such a step might involve.

To the core of the idea of relativizing the notion of quantum subsystems resides our discussion about emergence from Sec. \ref{sec:emergence} and Sec. \ref{sec:ontologies}, as there is a difference between our classical systems, including our lab devices and their single measurement outcomes, and their quantum descriptions as quantum subsystems. While it is their classical descriptions that are anchored in spacetime and invariant over time, their quantum representations can move freely within the overall Hilbert space and need to be re-established at every stage of the global Hilbert space evolution. Below, I sketch how this could work, consider the close resemblance between this approach and the methods used in general relativity to establish a local inertial frame, and speculate on connections to several other research fields.

\subsection{Quantum subsystems are not time-invariant}

Let's consider a global finite-dimensional (for simplicity) Hilbert space, $\Hcal$, the algebra of bounded operators on the Hilbert space, $B(\Hcal)$, a Hamiltonian $H$, and an initial pure state $\rho_0$. Then, for each observable $\Ocal \in B(\Hcal)$, we have
\begin{equation}
    v_{\Ocal}(\tau) = \tr \rho(\tau) \Ocal\,, \quad \text{where} \quad \rho(\tau) = e^{iH\tau}\rho_0e^{-iH\tau}\,, \label{eq:global_invs}
\end{equation}
where the unitary $U(\tau) \equiv e^{iH\tau}$ induces an inner automorphism of the algebra by the self-adjoint Hamiltonian and parametrized by the real parameter\footnote{Although we typically consider the evolution parameter $\tau$ to be time, particularly in light of the experimental success of applying Schrödinger's equation in lab, at this level it is just a parametrization of this inner automorphism. It is part of the work in establishing precisely the quantum-classical limit that the relationship between $\tau$ and geometric time $t$ should be stipulated. From this point of view, quantum mechanics is inherently background independent.} $\tau$. For any fixed $\tau_0$, the set $\{v_\Ocal\}_{\tau_0}$ fully parametrizes all the data of the theory, as the theory evolves deterministically. Note that $\{v_\Ocal\}_{\tau_0}$ is invariant under the following unitary map: 
\begin{equation}
    T:\Hcal \longrightarrow \bigotimes_I \Hcal^I\,,
\end{equation}
that transforms states and operators as: 
\begin{align}
    \Ocal &\longrightarrow T \Ocal T^{-1} \\
    |\Psi \rangle &\longrightarrow T |\Psi \rangle  \,,
\end{align}
which is the map used to define the tensor product structure in Definition \ref{def:TPS}. 

Let's now consider a subalgebra $\emph{App} \in B(\Hcal)$ associated with our experimental devices: $\Ocal_I^{\rm app}\in \emph{App}$, where  $\{\Ocal_I^{\rm app}\}$ form an orthogonal basis with $d^2-1$ elements for the apparatus. Moreover, let's assume, for now, that indeed our apparatuses are independent quantum systems (microcausality holds\footnote{The split property would be necessary around the region where our devices are located had we considered infinite-dimensional Hilbert spaces \cite{Fewster:2016mzz,EmerGe_proj_2}.}) such that $B(\Hcal)$ is isomorphic to $ B(\Hcal^{\rm app}) \otimes B(\Hcal^{\rm EE})$ (akin to \emph{completeness} in Sec. \ref{sec:sapce_QM} but now only for bounded operators), where EE stands for everything else, $\text{dim } \Hcal^{\rm app} =d$ and $\text{dim } \Hcal^{\rm EE} =D$, and $d/D \lll 1$. This is good since we typically do not have access to the whole Hilbert space, so we want to single out our apparatus. This implies that a TPS $\Tcal$ is being considered, such that:  
\begin{align}
    \Hcal &\overset{\Tcal}{\longrightarrow} \Hcal^{\rm app} \otimes \Hcal^{\rm EE} \\
    \Ocal^{\rm app}_I &\overset{\Tcal}{\longrightarrow} \tilde{O}^{\rm app}_I \otimes \mathds{I}^{\rm EE} \\ 
    \rho &\overset{\Tcal}{\longrightarrow} \sum_{n,m}^d \sqrt{p_n p_m} |u_n^{\rm app}\rangle \langle u_m^{\rm app}| v_n^{\rm EE}\rangle \langle v_m^{\rm EE}|\,,
\end{align}
where $\tilde{O}^{\rm app}_I\in B(\Hcal^{\rm app})$ and we used the Schmidt decomposition for a generic state $\rho$, so $\{|u_n^{\rm app} \rangle\}$ and $\{|v_n^{\rm EE} \rangle\}$ form orthonormal bases for $\Hcal^{\rm app}$ and $\Hcal^{EE}$, respec., and $p_n>0$. The Hamiltonian also factorizes: 
 \begin{equation}
     H \overset{\Tcal}{\longrightarrow} a_0 \mathds{I}_{d+D} + \sum_{I=1}^{d^2-1}a_I \tilde{\Ocal}^{\rm app}_I + \sum_{J=1}^{D^2-1}b_J \tilde{\Ocal}^{\rm EE}_J + \sum_{I=1}^{d^2-1}\sum_{J=1}^{D^2-1} c_{IJ} \tilde{\Ocal}^{\rm app}_I \tilde{\Ocal}^{\rm EE}_J\,,
 \end{equation}

Crucially, note that the quartet $(\Hcal,B(\Hcal), H,\rho_0)$ is now factorized, and associated with the factorization $\Tcal$ are the coefficients,
\begin{equation}
    \{p_n,a_0,a_I,b_J,c_{IJ}\}=\{p_n (\Tcal),a_0 (\Tcal),a_I (\Tcal),b_J (\Tcal),c_{IJ} (\Tcal)\} \,. \label{eq:coeeficients}
\end{equation}
which parametrize the states and the Hamiltonian, thus the kinematics and dynamics of the theory, in a given TPS. Their dependence on the TPS can be simply understood by the fact that our starting point did not have any of these coefficients present, thus they are contingent on the TPS chosen. On the other hand, they are not all independent from one another, as they are all connected to the class of unitary maps $\Tcal$ being considered, and its unitary nature will impose constraints on these parameters.

Under a given TPS, we can rewrite the global invariants \eqref{eq:global_invs} as
\begin{align}\label{eq:global_inv_convex}
    v_{\Ocal^{\rm app}_I} &= \sum_{m,n=1}^d \sqrt{p_n p_m} \tr_{\rm app} \left( \tilde{\Ocal}^{\rm app}_I |u_n^{\rm app} \rangle \langle u_m^{\rm app} | \right)  \tr_{\rm EE} \left(  |v_n^{\rm EE} \rangle \langle v_m^{\rm EE} | \right) \nonumber \\
    & =\sum_{n=1}^d p_n \tr_{\rm app} \left( \tilde{\Ocal}^{\rm app}_I P_n^{\rm app} \right) \,, \quad \text{where} \quad P_n^{\rm app} := |u_n^{\rm app} \rangle \langle u_n^{\rm app} |  \nonumber\\
    &= \sum_{n=1}^d p_n \tilde{v}_{\Ocal^{\rm app}_{I,n}} \,,
\end{align}
where $\tilde{v}_{\Ocal^{\rm app}_{I,n}}$ are the local (in the local unitary sense) invariants associated with the local observable $\tilde{O}_I^{\rm app}$ in the state $|u_n^{\rm app} \rangle$. Thus, generically, global invariants can be decomposed as a convex sum of local invariants. 


Now we introduce three crucial assumptions that will need to be further investigated in the future. The first two are strictly mathematical and demand a much closer study of TPS transformations, while the third one is based on our current understanding of LEQG, and quantum gravity more generally. It hints at the existence of a new dynamical equation in quantum mechanics that would be responsible for the evolution of the TPS in quantum systems, very much akin to the evolution of the metric in general relativity. They are:

\begin{itemize}
    \item For $d/D \lll1$ (which is absolutely the case), there are infinitely many inequivalent TPSs $\Tcal_i$ \cite{Stoica:2021rqi}. In fact, as they are connected by global unitaries that can be continuously parametrized, we will assume any two TPS are related by continuous deformations and parametrized by $\tau$: $\Tcal(\tau)$\footnote{Despite introducing this infinity to parameterize all the different TPS that fix the dimensionality of the Hilbert spaces of the apparatus and EE, it pales in comparison with all the possible different TPS that can be generically considered, e.g., bipartite with the same dimensionality for each subsystem or non-bipartite ones. Another way of understanding this is by taking the Hamiltonian perspective from \cite{Cotler:2017abq}, where one always considers a fixed TPS but then varies the Hamiltonian, akin to the coefficients in \eqref{eq:coeeficients} evolving. This is akin to how the renormalization group flow changes the values of the coupling parameters modeling interactions as the energy scale changes in the effective field theory paradigm, now understood as the notion of the relevant quantum subsystems being updated depending on which modes are integrated out.}; 

    \item There is a set of TPS, $\{\Tcal_M\}$, where the state of the system factorizes: $\rho \overset{\Tcal_M}{\longrightarrow} \rho^{\rm app'} \otimes \rho^{\rm EE'}$, where $'$ indicates that whatever quantity being considered is in such a TPS;

    \item Due to violations of microcausality in quantum gravity, imposing a fixed TPS in any quantum-gravitational system is ill-posed and can only be considered an approximation. As the system evolves under the global Hamiltonian, the approximate TPS changes continuously and can be parametrized by time, $\Tcal(\tau)$. This can be understood in terms of the coefficients above evolving, e.g., $a_I = a_I (\tau)$;

\end{itemize}
We will show that by putting together these assumptions, we can arrive at the following conclusion: 

\vspace{0.2cm}

\textit{ Single outcomes in any measurement correspond to a composition of a change of TPS together with a local unitary transformation of the local algebra assigned to the apparatus.}

\vspace{0.2cm}

\noindent In other words, a necessary condition for a single-outcome measurement at a given $\tau_0$ happens whenever $\Tcal(\tau_0) \in \{\Tcal_M\}$.

Let's start by showing this, and then we will discuss the assumptions and conclusion in light of a close analogy to finding local inertial frames in general relativity. Whenever $\Tcal(\tau_0) \in \{\Tcal_M\}$, we have a product state between what we call the apparatus and everything else, so we can write the state as 
\begin{equation}
    |\Psi (\tau_0)  \rangle  \overset{\Tcal_M}{\longrightarrow} \sum_{n=1}^{d} c_n |u_n^{\rm app'} \rangle \otimes \sum_{m=1}^D d_m |v_m^{\rm EE'} \rangle\,,
\end{equation}
where these are bases for the apparatus and everything else in one of the $\Tcal_M$ (to not get confused with the bases introduced in \eqref{eq:global_inv_convex}). Now, we can simply \emph{rotate/align} our apparatus by a local unitary transformation, such that
\begin{equation}
    |\Psi (\tau_0)  \rangle =  |o_{I,n}^{\rm app'} \rangle \otimes \sum_m^D d_m |v_m^{\rm EE'}\rangle \, \quad \text{where} \quad \bar{\Ocal}^{\rm app'}_I  |o_{I,n}^{\rm app'} \rangle = o_{I,n}^{\rm app'} |o_{I,n}^{\rm app'} \rangle\,,
\end{equation}
and $\bar{\Ocal}^{\rm app'}_I= U^{-1}\tilde{\Ocal}^{\rm app}_IU$. Thus,  the local state is one of the rotated apparatus' eigenstates with eigenvalue $o_{I,n}^{\rm app'}$. Crucially, we do not know which one. Then, it is easy to show that, 
\begin{align}
    v_{\Ocal^{\rm app}_I} &= \tr_{\rm app'} \left(|o_{I,n}^{\rm app'} \rangle \langle o_{I,n}^{\rm app'} |\bar{\Ocal}^{\rm app'}_I \right) \tr_{\rm EE'} \rho^{\rm EE'} \nonumber \\
    & = o_{I,n}^{\rm app'} \,.
\end{align}

Therefore, by identifying the appropriate TPS and locally rotating/aligning the apparatus, we see that local measurement outcomes are actually the global invariants that we started with, closing the gap between the holistic and atomistic ontologies in quantum mechanics at the empirical level. Crucially, to arrive at this point, besides our assumptions (to be scrutinized), we only relied on global and local unitary transformations. We expect that the probabilistic nature of the theory arises from our lack of knowledge about the specific set of unitaries required to reach this point, such that probabilities will be epistemic. This will be closely investigated in the future. In any case, given that all considerations involved only unitary transformations (both for TPS updates and time evolution) and we still arrive at a description of the apparatus with single outcomes related to the global invariants, we term this yet-to-be-finalized theory \emph{single-world unitary quantum mechanics}. Under $\mathcal{T}_M$, local measurements can be associated with single outcomes, even though all transformations considered were unitary.

\subsection{Comparison with general relativity}

Let's clarify the preceding discussion by drawing a comparison with a similar procedure in general relativity. As one moves through spacetime and aims to consistently compare measurements at different points along their trajectory, it is essential to update their local inertial frames (tetrads) based on the local metric of spacetime. To achieve this, one must first solve the Einstein field equations for the metric, which informs the local geometry and allows for the determination of tetrads. Tetrads provide local orthonormal bases, with the timelike tetrad vector typically aligned with the observer’s 4-velocity to serve as the time direction in the observer’s local frame. If the timelike tetrad vector is not aligned with the observer’s 4-velocity, it represents a boosted frame relative to the observer’s rest frame, leading to different local measurements of time and space, akin to special relativity. Therefore, it is crucial for the observer to measure their relative velocity in relation to some reference points to determine the appropriate tetrad to use.

The core process involves three key steps to define the appropriate local inertial frame: first, determine the spacetime metric by solving the Einstein field equations, which, despite being local equations, often require \textit{global} boundary conditions for a solution; second, establish the family of local inertial frames using the metric; and finally, align the timelike tetrad with the observer’s 4-velocity, which is a local process requiring measurements of the observer’s relative velocity to reference points.

Thus, the situation surrounding measurements in quantum mechanics seems quite analogous. The third assumption above hints towards an analog to the Einstein equations for the metric, but now a new dynamical equation for $\mathcal{T}(\tau)$, involving the global evolution of states. Once that is known, we can establish which different TPSs are temporarily available, which, like tetrads, need to be updated over time, allowing the introduction of approximate well-defined local quantum subsystems. Finally, one needs to make local measurements to \emph{orient} their local basis such that measurements yield single outcomes.

\subsection{Discussion}

Within this approach, we have directly connected insights from quantum gravity and its lack of gauge-invariant local observables to the measurement problem in quantum mechanics. Although much work needs to be done, if this program succeeds, it would mean that all classical measurements are interconnected by a web of measurement dualities. The discussion was limited to bipartite systems, but typically we also consider an observer in a quantum measurement, which nowadays is much better discussed and formalized in the context of quantum reference frames\footnote{There is a large literature on this topic \cite[e.g.]{Aharonov:1967zza, Aharonov:1984zz, Bartlett:2006tzx, Vanrietvelde:2018pgb, Giacomini:2017zju, delaHamette:2020dyi, Loveridge:2017pcv, Castro-Ruiz:2021vnq, delaHamette:2021oex, AliAhmad:2021adn, Lake:2023nua,AliAhmad:2024wja}. For a recent paper, see \cite[and references 1-23 therein]{Kabel:2024lzr}.}. In this program, changes of TPS appear naturally as one considers transformations between quantum systems to select the one that will be the reference frame, very similar to what we do in special relativity to choose the preferred inertial frame considered at rest. However, the status of the program remains mostly kinematical, in the sense that they do not consider the change of TPS due to time evolution for the same quantum system, similar to a classical frame having to update its local inertial frame as it moves in a dynamical spacetime. Three recent papers that delve into dynamics are \cite{Chandrasekaran:2022cip, Fewster:2024pur, DeVuyst:2024pop}. These works, together with the formalism introduced in \cite{Glowacki:2024jhi}, might allow for an extension of this proposal to infinite-dimensional Hilbert spaces and help better quantify the mathematical assumptions surrounding the existence of a large and continuous set of different TPS related to classical measurements.

%% file: inside.tex
\section{We are inside the box!}

The most vivid image you can retrieve from this whole story is that we are inside that initial mysterious box! In fact, due to the seemingly unavoidable non-local nature of gauge-invariant observables in quantum gravity, to some extent, we are the box. This is why holography has been one of the main entries into quantum gravity over the last few decades.

The issue we highlight here is that, generally, we are not using holography to solve elementary problems in physics involving gravity and quantum phenomena\footnote{As usual, there are counterexamples such as \cite[and references therein]{Nojiri_2019,Nastase_2020}.}. Typically, we rely on perturbative quantum gravity under standard covariant quantization. Despite preventing us from establishing gauge-invariant local observables, we write down models in these contexts as if that were not an issue. And, in truth, it works—to the extent that we can conduct experiments in our labs and explain them successfully. Naturally, this raises two questions: why does it work, and where should we expect it not to work?

In \cite{Donnelly:2017jcd}, an answer to the first question is proposed. For classical gravity, they demonstrate that certain classical configurations can be gravitationally dressed in a way that the dressing depends only on the Poincaré charges (energy, momentum, and angular momentum) of the configuration. Different configurations with the same Poincaré charges are indistinguishable from outside their region, suggesting a way to localize information classically. In the quantum context, they argue that perturbative quantum gravity in the presence of global symmetries allows for the definition of localized information or ``gravitational qubits," which can be localized to regions in such a way that their information content is not accessible outside the region. Meanwhile, \cite{Goeller:2022rsx} seems to obtain a general and fully non-perturbative framework for constructing dynamical reference frames in generally covariant theories. By constructing relationally local observables throughout spacetime that satisfy microcausality in the bulk, they establish what they call ``relational locality.''

In our attempt to answer the second question, we propose further scrutiny of early universe models that use perturbative quantum gravity to explain the origin of cosmological perturbations and the modeling of gravitational-induced-entanglement experiments. We also implied that the measurement problem might be a consequence of the global imposition of a TPS under unitary evolution. In fact, within our approach, there also seems to exist some notion of relational locality, as each tensor product structure temporarily induces a notion of microcausality that is continuously updated under global unitary evolution. It will be interesting to see whether their work can provide the mathematical framework in which single-world quantum mechanics can be embedded.

We end by observing that progress is typically made when obvious notions are finally put into question. Among the historically remarkable examples, perhaps Einstein's scrutinizing of the nature of simultaneity leading to the theory of relativity and Planck's debunking of the radiation energy spectrum continuum giving rise to quantum mechanics stand out the most. Another seemingly obvious concept is the idea of independent physical systems. Combining relativity and quantum mechanics challenges the immediate naive understanding of how to independently describe different interacting physical systems, especially across different levels of descriptions, so we hope that further analyzing this notion will pave the way towards a better understanding of quantum mechanics and quantum gravity.

%% file: acknowledgements.tex
This work is the result of numerous discussions with many people over the last two years. I am deeply grateful to the \href{https://emerge-collab.org/}{EmerGe} (Emergent Geometries) collaboration, which includes Eugenia Colafranceschi, Andrea Di Biagio, Joakim Flinckman, Jan Głowacki, Niels Linnemann, and Florian Niedermann. Collaborating with them has allowed me to deepen my knowledge and interest in classical and quantum subsystems. This interest began during the Sejny Summer Institute in 2021, sparked by conversations with Pierre Martin-Dussaud. I have discussed using changes in the TPS to address the measurement problem with Andrea Di Biagio, Flavio Del Santo, Patrick Fraser, Arsalan Adil, Leon Loveridge, Anne-Catherine de la Hamette, Sören Holst, and Jan Głowacki. The last three provided feedback on the discussions presented in this paper. I am also thankful to Vasileios Fragkos for his feedback on the section on GIE experiments, as well as to Benjamin Bose and Florian Niedermann for reviewing parts of the draft. I am also thankful to Emily Adlam, Tales Rick Perche and Shadi Ali Ahmad for careful reading and further comments on the draft. Finally, many ideas concerning the measurement problem emerged from supervising the master's thesis of Nicolas Boulles, with whom I had extensive discussions on the topic. This work was supported by the Swedish Research Council (grant number 2022-01893VR).